%% file: main.tex
\documentclass{article}

\PassOptionsToPackage{numbers, compress}{natbib}

\usepackage{tcolorbox}
\definecolor{cadmiumgreen}{rgb}{0.0, 0.42, 0.24}
\definecolor{oldmauve}{rgb}{0.4, 0.19, 0.28}
\definecolor{royalazure}{rgb}{0.0, 0.22, 0.66}
\definecolor{harvardcrimson}{rgb}{0.79, 0.0, 0.09}
\definecolor{lightmauve}{rgb}{0.86, 0.82, 1.0}
\definecolor{darkbrown}{rgb}{0.4, 0.26, 0.13}%
\usepackage[pagebackref=true,breaklinks=true,colorlinks,bookmarks=false]{hyperref}

\usepackage[final]{neurips_2024}

\usepackage[utf8]{inputenc} 
\usepackage[T1]{fontenc}    
\usepackage{hyperref}       
\usepackage{url}            
\usepackage{booktabs}       
\usepackage{amsfonts}       
\usepackage{nicefrac}       
\usepackage{microtype}      
\usepackage{xcolor}         
\usepackage{algorithm}
\usepackage{algorithmic}
\usepackage{float}
\usepackage{booktabs}
\usepackage{multirow}
\usepackage{capt-of}
\usepackage{wrapfig,lipsum,booktabs}

\usepackage{enumitem,graphicx}
\input{pkgs/math_commands}

\title{Mitigating Backdoor Attack by Injecting Proactive Defensive Backdoor}

\newcommand*{\email}[1]{%
    \normalsize\href{mailto:#1}{#1}\par
    }
\author{%
\ \ \ \ Shaokui Wei\textsuperscript{1}
\ \ \ \ Hongyuan Zha\textsuperscript{1,2}
\ \ \ \ Baoyuan Wu\textsuperscript{1}\thanks{Corresponds to Baoyuan Wu (\email{wubaoyuan@cuhk.edu.cn}).} 
\\
\textsuperscript{1}School of Data Science, \\The Chinese University of Hong Kong, Shenzhen, Guangdong, 518172, P.R. China\\
\textsuperscript{2}Shenzhen Key Laboratory of Crowd Intelligence Empowered Low-Carbon Energy Network
 }

\def\Dtr{\mathcal{D}_{tr}}
\def\Dcl{\mathcal{D}_{cl}}

\def\Ddef{\hat{\gD}_{def}}

\def\app{\textbf{Appendix}}

\begin{document}

\maketitle

\begin{abstract}
Data-poisoning backdoor attacks are serious security threats to machine learning models, where an adversary can manipulate the training dataset to inject backdoors into models. In this paper, we focus on in-training backdoor defense, aiming to train a clean model even when the dataset may be potentially poisoned. Unlike most existing methods that primarily detect and remove/unlearn suspicious samples to mitigate malicious backdoor attacks, we propose a novel defense approach called PDB (\textbf{P}roactive \textbf{D}efensive \textbf{B}ackdoor). Specifically, PDB leverages the “home field” advantage of defenders by proactively injecting a defensive backdoor into the model during training. Taking advantage of controlling the training process, the defensive backdoor is designed to suppress the malicious backdoor effectively while remaining secret to attackers. In addition, we introduce a reversible mapping to determine the defensive target label. During inference, PDB embeds a defensive trigger in the inputs and reverses the model’s prediction, suppressing malicious backdoor and ensuring the model's utility on the original task. Experimental results across various datasets and models demonstrate that our approach achieves state-of-the-art defense performance against a wide range of backdoor attacks. The code is available at \href{https://github.com/shawkui/Proactive_Defensive_Backdoor}{https://github.com/shawkui/Proactive\_Defensive\_Backdoor}.
\end{abstract}

\section{Introduction}

\input{section/introduction}

\section{Related work}

\input{section/related_work}

\section{Method}

\input{section/methods}

\section{Experiments}
\label{main_exp}

\input{section/experiment}

\section{Conclusion}

\input{section/conclusion}

\begin{ack}
Baoyuan Wu is supported by Guangdong Basic and Applied Basic Research Foundation (No. 2024B1515020095), National Natural Science Foundation of China (No. 62076213), Shenzhen Science and Technology Program under grants (No. RCYX20210609103057050), and Longgang District Key Laboratory of Intelligent Digital Economy Security. Hongyuan Zha is supported in part by the Shenzhen Key Lab of Crowd Intelligence Empowered Low-Carbon Energy Network (No. ZDSYS20220606100601002). This work is supported by Shenzhen Science and Technology Program under grant No. GXWD20201231105722002-20200901175001001, and No. ZDSYS20211021111415025, and No. JCYJ20210324120011032, and the Guangdong Provincial Key Laboratory of Big Data Computing, the Chinese University of Hong Kong, Shenzhen. 
\end{ack}

\newpage
\bibliographystyle{plainnat}
\bibliography{references}

\clearpage
\appendix

\input{appendix/exp_detail}

\input{appendix/exp_add}

\input{appendix/add_analysis}

\clearpage
\input{section/checklist}

\end{document}

%% file: pkgs/math_commands.tex
\usepackage{amsmath,amsfonts,bm}

\def\ie{$i.e.$}
\def\eg{$e.g.$}

\newcommand{\wrt}{\textit{w.r.t.~}}

\def\1{\bm{1}}

\def\vtheta{{\bm{\theta}}}

\def\vx{{\bm{x}}}

\def\veps{{\bm{\epsilon}}}

\def\gD{{\mathcal{D}}}

\def\gX{{\mathcal{X}}}
\def\gY{{\mathcal{Y}}}



%% file: section/introduction.tex
In recent years, deep neural networks (DNNs) have become ubiquitous across diverse fields, powering applications such as face recognition, self-driving vehicles, and medical image analysis \citep{adjabi2020past, he2016deep, liu2020computing, tournier2019mrtrix3}. However, alongside these advancements, the vulnerability of DNNs to malicious attacks presents a critical challenge to their safety and reliability. A particularly alarming threat arises from backdoor attacks, where adversaries secretly introduce backdoors into DNN models during training by subtly altering a fraction of the dataset. This manipulation ensures the model's standard performance on uncontaminated data but erroneously assigns a pre-determined label to any input carrying a specific trigger. Considering its real threats to machine learning systems, especially in security-critical scenarios, it’s a practical necessity to investigate and propose effective defense strategies against such attacks to safeguard real-world applications.

To mitigate the threats posed by backdoor attacks, researchers have actively explored various backdoor defense techniques throughout the life cycle of machine learning systems \cite{wu2023defenses}. In this paper, we specifically delve into  \textbf{in-training backdoor defense} \cite{wubackdoorbench, wu2023defenses, wu2024backdoorbench}, which aims to train machine learning models using datasets that may be contaminated with poisoned data.
Most existing methods in this field primarily focuses on identifying suspicious samples through various means, along with mitigating the backdoor effect by directly removing \cite{chen2019detecting, yuan2023activation} or applying some techniques (\eg, unlearning \cite{li2021anti,chen2022effective}, or relabel \cite{huang2022backdoor, liu2023beating, zhu2023victim}) to the suspicious samples. 
Despite achieving remarkable performance in backdoor defense, these methods face certain limitations and challenges. First, most existing works rely on specific assumptions such as the latent separability \cite{chen2019detecting} or the early learning of poisoned samples \cite{li2021anti, zhu2023victim, zhang2023backdoor} to identify the poisoned samples. However, these assumptions may not hold under more sophisticated attacks \cite{qi2023revisiting}. As accurately detecting poisoned samples is crucial for those methods, any deviation from their underlying assumptions could lead to performance degradation and compromise their effectiveness. Second, some methods, such as DBD \cite{huang2022backdoor}, NAB \cite{liu2023beating}, and V\&B \cite{zhu2023victim}, necessitate complex modifications to the training process, resulting in a substantial increase in training costs.

In this paper, instead of following the traditional detection-and-mitigation pipeline, we propose a proactive approach that leverages the “\textit{home field}” advantage of defenders. Our method, called PDB (short for \textbf{P}roactive \textbf{D}efensive \textbf{B}ackdoor), aims to fight malicious backdoor attacks by injecting a proactive defensive backdoor introduced by the defenders themselves. The primary objective of PDB is to suppress the malicious backdoor with a defensive backdoor while keeping the model's utility on original task. Specifically, When the defensive trigger is presented, the defensive backdoor will dominate the prediction of the proactively backdoored model, effectively suppressing the malicious backdoor’s impact. Importantly, our defensive backdoor allows for the restoration of the ground truth label to maintain the model’s utility on the original task. To achieve this goal, we first analyze the objective for an effective defensive backdoor and introduce four essential design principles, including \textbf{reversibility}, \textbf{inaccessibility to attackers}, \textbf{minimal impact on model performance}, and \textbf{resistance against other backdoors}. Then, we construct an additional defensive poisoned dataset, subsequently utilizing such dataset and the whole poisoned dataset to train the model.  Consequently, if only the malicious trigger is present, the model remains under the control of the malicious backdoor. However, when the defensive trigger appears, the defensive backdoor is activated, mitigating the malicious backdoor effect. To evaluate its effectiveness, we compare PDB with five state-of-the-art (SOTA) in-training defense methods across seven SOTA data-poisoning backdoor attack methods involving different model structures and datasets. Our experimental results demonstrate that PDB achieves comparable or even superior performance compared to existing baselines.

Our main contributions are threefold: 
\textbf{1)} We break away from the traditional detection-and-mitigation pipeline by proposing a novel mechanism that injects a proactive defensive backdoor during training, which suppresses the malicious backdoor while preserving the model's utility on the original task, without any specific assumptions about potential malicious backdoor attacks.
\textbf{2)} By analyzing the primary objective, we introduce essential design principles for an effective defensive backdoor and propose a practical algorithm to implement the defensive backdoor. 
\textbf{3)} We conduct extensive experiments to evaluate the effectiveness of our method and compare it with five SOTA defense methods across seven challenging backdoor attacks, spanning diverse model structures and datasets, demonstrating the superior performance of the proposed method.

%% file: section/related_work.tex
\paragraph{Backdoor attacks.}

DNNs face significant security threats from backdoor attacks, which are designed to maintain normal performance on regular inputs while forcing the network to output a predetermined target when a specific trigger is introduced. These attacks can be generally categorized into two types based on the property of the trigger: static-pattern backdoor attacks and dynamic-pattern backdoor attacks. The seminal instance of static-pattern backdoors, BadNets \cite{gu2019badnets}, employed fixed triggers like white squares. To enhance trigger stealthness, the Blended approach \cite{chen2017targeted} was introduced, which merges the trigger with the host image in a subtle manner. Recognizing the potential for detection in fixed-pattern triggers, the research has pivoted towards dynamic-pattern backdoor attacks. Innovations in this direction, such as SSBA \cite{li2021invisible}, WaNet \cite{nguyen2021wanet}, LF \cite{zeng2021rethinking}, WPDA \cite{song2024wpda},  IRBA \cite{gao2023imperceptible}, VSSC \cite{wang2023robust} and TAT \cite{cheng2023tat}, have focused on crafting sample-specific triggers that are more challenging to identify. Techniques to refine the stealthness of triggers have been furthered by works like Sleeper-agent \cite{souri2022sleeper} and Lira \cite{doan2021lira}, which optimize the output to be more covert. The sophistication of backdoor attacks has recently been advanced by strategies for learning-based poisoning sample selection \cite{zi2023boost} and re-activation attack \cite{zhu2024breaking}. To execute attacks without altering the consistency between the image and its label, 'clean label' attacks have been introduced. For example, LC \cite{shafahi2018poison} and SIG \cite{barni2019new} employed counterfactual methods and additional techniques to modify the image while maintaining label consistency subtly.

\paragraph{Backdoor defenses.}

The main purpose of backdoor defense is to alleviate the vulnerabilities of DNNs to backdoor attacks by employing various strategies during different stages of the model lifecycle. Therefore, backdoor defenses are typically categorized into three types: pre-training, in-training, and post-training. Pre-training defenses concentrate on the detection and removal of poisoned data points before training. For example, AC \cite{chen2019detecting} leverages unusual activation patterns to weed out poisoned data, while Confusion Training \cite{qi2023towards} relies on a model trained specifically to recognize poisoned instances. VDC \cite{zhu2023vdc} utilizes the capabilities of large multimodal language models for the same purpose. 
Post-training defenses are applied after a model has been trained. A line of works in this direction focusing on pruning \cite{liu2018fine,wu2021adversarial, zheng2022preactivation, zheng2022data, lin2024unveiling} or fine-tuning \cite{Zhu_2023_ICCV,min2024towards} to neutralize the backdoor. Besides,  I-BAU \cite{zeng2022adversarial}, NPD \cite{zhu2023neural}, and SAU \cite{wei2024shared} reverse potential backdoor triggers by adversarial techniques to cleanse the model. NAD \cite{li2021neural} employs a slightly poisoned model to assist in retraining a heavily compromised one.

This paper mainly focuses on the in-training defenses that aim to prevent backdoor insertion during the training phase. Along this direction, ABL \cite{li2021anti} utilizes the observation that the poisoned samples are easier to learn than normal samples, resulting in the different learning speeds between benign and poisoned samples, to detect and unlearn the poisoned samples. Based on similar observation, V\&B \cite{zhu2023victim} first trains a backdoored model to capture the backdoor effect and utilizes the backdoored model to train a benign model by detecting and applying a series of operations on the suspicious samples. Similarly, CBD \cite{zhang2023backdoor} first trains a backdoored model for a few epochs and trains a benign model by reweighting the samples and deconfounding the representation. DBD \cite{huang2022backdoor} splits the training process into three steps and employs self-supervised learning techniques to detect suspicious samples and train a benign model. D-ST \cite{chen2022effective} leverages the fact that benign samples are less sensitive to image transformations to detect suspicious samples and employs semi-supervised learning to train a benign model. Recently, a few attempts have been made to defend against malicious attacks by incorporating proactive attacks \cite{zhu2023ai, liu2023beating}. The work most closely aligned with our approach is NAB \cite{liu2023beating}, which first identifies and then relabels potentially poisoned samples in the dataset, subsequently embedding non-adversarial triggers into the suspicious samples to mitigate the backdoor effect. In contrast to their methodology, our technique offers a more straightforward solution, eliminating the need for costly detection and relabeling processes, thus reducing overall costs and complexity. In essence, we demonstrate that injecting a defensive backdoor alone is sufficient to defend against backdoor attacks without requiring detection and relabeling of the poisoned samples. We refer readers to \cite{wu2023defenses} for more defense in adversarial machine learning.

%% file: section/methods.tex
In Section~\ref{sec::pre}, we introduce the essential notations and define the threat model in this paper. Subsequently, we explore the principles behind effective defensive backdoors, illustrated with practical examples in Section~\ref{sec::design}. We present the overall pipeline for our proposed method in Section~\ref{sec::fb3}.

\subsection{Problem setting}
\label{sec::pre}

\paragraph{Notations.} 
 Considering a sample $\vx \in \gX$ with label $y\in \gY$, a DNN model $f_\vtheta$ parameterized by $\vtheta$ is trained to classify $\vx$.
 The space $\gY=[1, \cdots, K]$ denotes the space of candidate labels ($K\geq 2$), and $\gX$ represents the sample space. 
 In the context of backdoor attack, we denote the trigger by $\Delta$ and the trigger injection operator by $\oplus$. Consequently, given a benign sample $\vx$, the poisoned sample can be generated by $\vx\oplus\Delta$. It's important to note that the injection operator $\oplus$ can vary according to the type of trigger $\Delta$.

\paragraph{Threat model.}
We consider a data poisoning scenario for \textbf{malicious backdoor} attack where the attacker can only manipulate a portion of the training dataset to plant trigge but cannot control the training process. By poisoning the dataset, the model trained on the manipulated dataset $\Dtr$ would normally perform for benign input but classify the inputs with malicious trigger $\Delta$ to predefined target $\hat{y}$. Besides, we define the portion of manipulated samples as the \textbf{poisoning ratio} of backdoor attack. 

 The defender faces a situation where a potentially poisoned dataset is given. The defender aims to train a model where the malicious backdoor fails to be activated by the malicious trigger, and the model's utility on the original task is maintained. We assume a small benign dataset $\Dcl$ is reserved for the defender, which can be obtained by various means, including but not limited to purchase from reputable data vendors, generation via state-of-the-art generative models \cite{croitoru2023diffusion, goodfellow2020generative, kingma2013auto}, collection from the internet, or applying data cleansing methods \cite{wu2023defenses}. Moreover, we assume that the defender does not have knowledge of either the malicious trigger $\Delta$ or the malicious target label $\hat{y}$.

\subsection{Proactive defensive backdoor }
\label{sec::design}
In this paper, we aim to defend the unknown malicious backdoor with the trigger $\Delta$, by inserting a proactive defensive backdoor with a trigger $\Delta_1$ into the model.
Our primary objective is to ensure that when $\Delta_1$ is presented, the model's output will be controlled by $\Delta_1$ rather than $\Delta$, thereby suppressing the malicious backdoor. Besides, the model's utility on the original task should be preserved, \ie, user can still get the true prediction of the benign sample with the defensive trigger.  To achieve such a defense goal, the desired defensive backdoor attack should follow the principles below:

\begin{itemize}[leftmargin=*]
    \item \textbf{Principle 1: Reversibility.} The defensive backdoor must be reversible, such that the ground truth label can be restored from the prediction of benign samples attached with $\Delta_1$. Such a requirement is crucial for preserving the model performance on benign inputs with $\Delta_1$.
    
    \item \textbf{Principle 2: Inaccessibility to attackers.} The defensive trigger $\Delta_1$ should be meticulously designed to be non-replicable and undiscoverable by potential attackers. By doing so, we prevent adversaries from exploiting the same trigger or using inversion techniques to identify it.
    
    \item \textbf{Principle 3: Minimal impact on model performance.} While stealth is not a strict requirement for the defensive trigger, modified samples should retain sufficient characteristics of the original data. This ensures accurate label recovery from the model’s predictions in the presence of $\Delta_1$.

    \item \textbf{Principle 4: Resistance against other backdoors.} To effectively mitigate malicious backdoors, the defensive backdoor should be resistant to various backdoor attacks, not only known attacks but also potential future backdoors.
\end{itemize}

In light of the principles outlined above, we delve into the practical design of our defensive backdoor\footnote{It’s important to acknowledge that alternative designs may also adhere to these principles.}. 

\begin{wrapfigure}{r}{0.5\textwidth}
  \vspace{-0.25in}
  \begin{center}
    \includegraphics[width=0.5\textwidth]{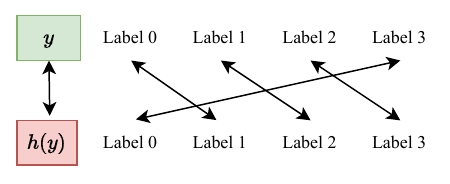}
  \end{center}
  \vspace{-0.15in}
  \caption{Illustration of bijective mapping with $h(y)=(y+1)\mod K$, with $K=4$.}
  \label{map}
  \vspace{-0.1in}
\end{wrapfigure}

\paragraph{Following Principle 1.} For the first principle, we propose to assign the target label by a bijective mapping 
 $h:\gY\to\gY$, such that the target label of a sample with label $y$ is $h(y)$ and the ground truth label of a poisoned image with label $y$ is $h^{-1}(y)$. A typical choice of $h$ and $h^{-1}$ is $h(y)=(y+1)\mod K$ and $h^{-1}(y)=(y-1)\mod K$ where $\mod$ represents the modulo operation and $K$ is the number of classes. It's worth noting that in the context of DNNs, $h$ can also be formulated as a function of logits or features such as $h(\phi(\vx))=-\phi(\vx)$ and  $h^{-1}(\phi(\vx))=-\phi(\vx)$ where $\phi(\vx)$ corresponds to the features or logits of input. This flexibility allows for a broader range of target label assignment strategies.

\begin{wrapfigure}{r}{0.5\textwidth}
  \vspace{-0.27in}
  \begin{center}
    \includegraphics[width=0.5\textwidth]{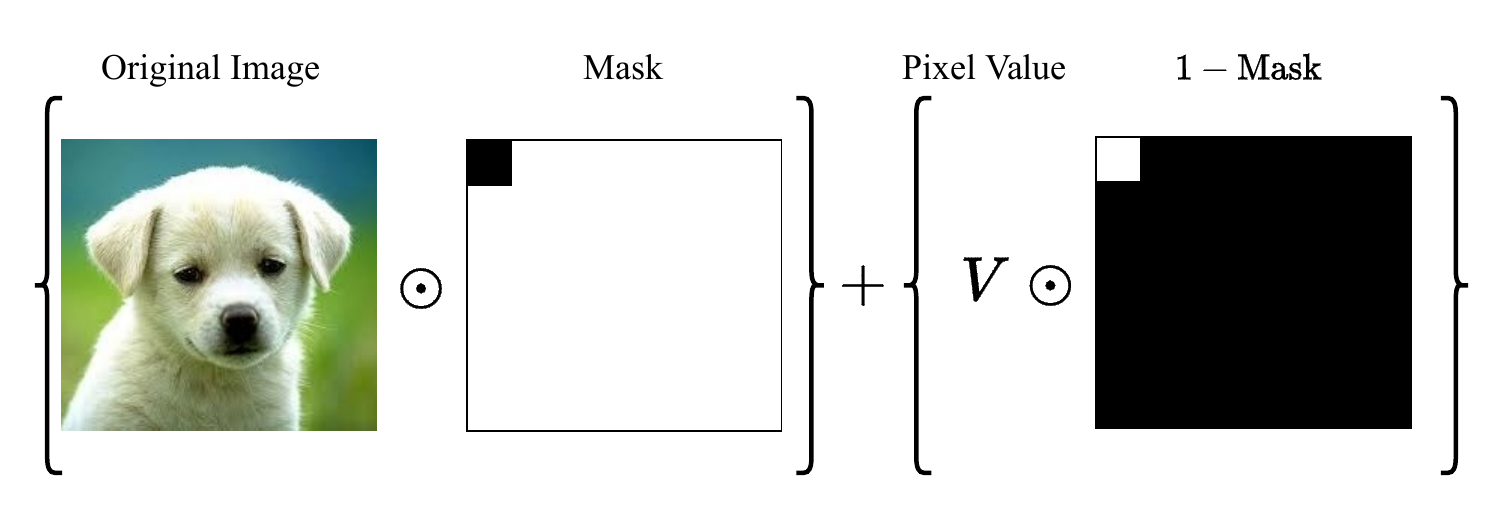}
  \end{center}
  \vspace{-0.15in}
  \caption{Demonstration of generating a defensive poisoned sample. $V\notin[0,1]$ is the pixel value of trigger, $\odot$ is the element-wise product. For the mask, $0$ is represented by black, while $1$ is represented by white.}
  \label{pi}
  \vspace{-0.2in}
\end{wrapfigure}

\paragraph{Following Principle 2 \& 3.} To follow the second and third principles, the design of the trigger is essential. Consider the patched trigger as an illustrative example, which can be constructed by carefully determining its position and pattern. Regarding the trigger’s position, it should be crafted to preserve the core visual patterns of the original image, ensuring that the primary content remains unaltered. As for the trigger’s pattern, we leverage the \emph{“home field”} advantage of the defender, designing a trigger that operates beyond the conventional pixel space. Specifically, for an image with pixel values in the range of $[0,1]$
, the trigger is engineered to modify regions to values beyond this range. This modification renders the trigger infeasible and not invertible by attackers, given the natural constraints of image data.

\paragraph{Following Principle 4.} Following the fourth principle, the defensive backdoor is required to be resistant against other backdoors in the dataset. To meet such requirements, the key point is that the defender can control the training process, a \emph{"home filed"} advantage that attackers lack. On the one hand, the defender can design a strong defensive backdoor, \eg, adopting a large trigger. On the other hand, the defensive backdoor can be further enhanced by controlling the training process, \eg, applying data augmentation or adjusting the weight of defensive poisoned samples. More discussion and empirical findings are presented in \app~\ref{app:p4}.

\subsection{Backdoor injection}
\label{sec::fb3}

\begin{figure}[ht]
    \centering
    \includegraphics[width = \linewidth]{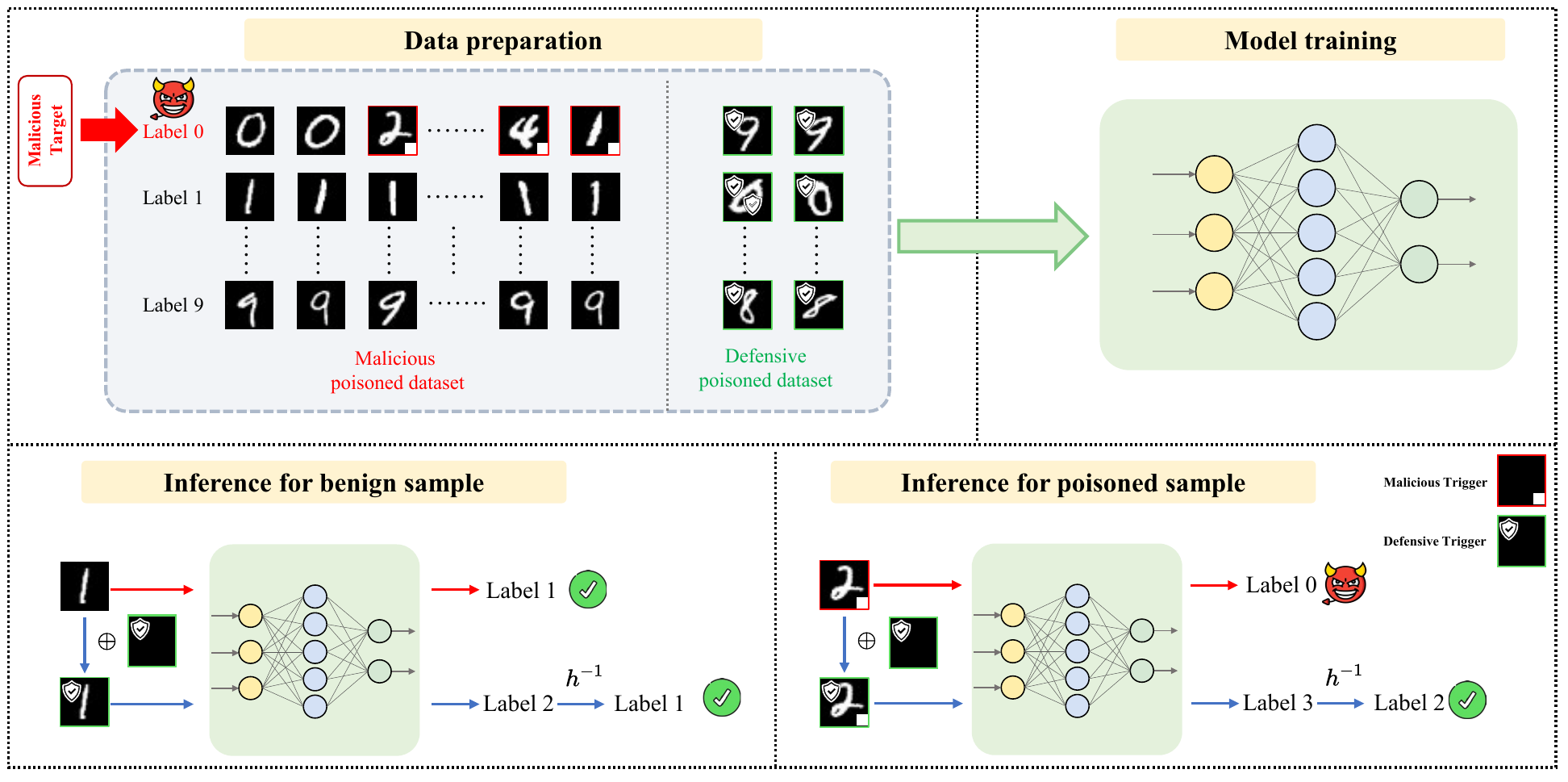}
    \caption{Overview of the proposed method. The trigger of the malicious backdoor is a white square, and its target label is $0$. The trigger of the defensive backdoor is represented by a white shield, and the target label mapping is $h(y)= (y+1)\mod 10$ and  $h^{-1}(y)= (y-1)\mod 10$ .}
    \label{fig::overview}
\end{figure}

As depicted in Figure~\ref{fig::overview}, our proposed method involves three key steps:

\paragraph{Data preparation.} Given a well-designed defensive backdoor with trigger $\Delta_1$ and a target label mapping $h$, a defensive poisoned dataset is first constructed by
\begin{equation}
     \Ddef= \{(\vx\oplus \Delta_1, h(y))|\forall (\vx,y) \in \Dcl\}.
\end{equation}

\paragraph{Model training.} Now, a model can be trained on the combination of the malicious poisoned dataset $\Dtr$ and the defensive poisoned dataset $\Ddef$. Then, a well-trained model will normally perform for benign inputs while controlled by the corresponding backdoor when either the trigger $\Delta$ or $\Delta_1$ is presented. However, if both $\Delta$ and $\Delta_1$ are simultaneously presented, the model may become confused due to the lack of such samples in the training dataset. As aforementioned, to ensure that the defensive trigger $\Delta_1$ effectively defeats an unknown trigger $\Delta$, some \emph{backdoor enhancement strategies} such as data augmentation or increasing sample weight can be adopted to enhance the defensive backdoor. In summary, the overall training objective is formulated as follows:
\begin{equation}
    \label{eq::1}
    \min_{\vtheta} \sum_{(\vx,y)\in \Dtr}L_0(f_{\vtheta}(\vx),y) + \sum_{(\vx,y)\in \Ddef} \lambda_1 L_1(f_{\vtheta}(\vx),y) + \lambda_2 L_2(f_{\vtheta}(\tau(\vx)),y), 
\end{equation}
where $\Dtr$ and $\Ddef$ are the maliciously poisoned training dataset and the defensive poisoned dataset, respectively. The operation $\tau$ enhances the defensive backdoor by applying operation on the defensive poisoned samples (\eg, adding noise: $\tau(\vx) = \vx + \veps$ with $\veps \sim \mathcal{N}(0, 1)$). 

In $(\ref{eq::1})$, the first term stands for the loss on the poisoned dataset, the second term stands for the loss of injecting our defensive backdoor, and the third loss aims to enhance the defensive backdoor. We use $L_0$, $L_1$, and $L_2$ to represent the loss function for each term, which are usually Cross-Entropy losses if not specified.  The parameters $\lambda_1$ and $\lambda_2$ are introduced to balance the contributions of the respective loss components. More details for the model training and implementation can be found in \app~\ref{app:exp_detail}.

\paragraph{Inference.}  During the inference, each input sample $\vx$ is initially embedded with the defensive trigger, and the model's prediction $f_{\vtheta}(\vx\oplus\Delta_1)$ is obtained. Subsequently, the authentic prediction is reconstructed via the inverse mapping $h^{-1}(f_{\vtheta}(\vx\oplus\Delta_1))$.

Below, we provide a high-level pseudocode representation of our proposed method for training and inference:

\begin{algorithm}[H]
\caption{Proactive Defensive Backdoor (PDB)\label{alg:FB3}}
\begin{algorithmic}
\STATE \textbf{Input:} Model $f_{\vtheta}$, poisoned training set $\Dtr$, reserved benign dataset $\Dcl$, defensive trigger $\Delta_1$, defensive target mapping $h$, max iteration number $T$. 
\STATE Initialize $f_{\vtheta}$.
\STATE $\triangleright$ \quad Data preparation
\STATE Construct the defensive poisoned dataset $\Ddef=\{(\vx\oplus\Delta_1, h(y)|(\vx,y)\in \Dcl)$.
\STATE $\triangleright$ \quad Model training
\FOR {$t=0,...,T-1$}
\FOR {each mini-batch in $\Dtr\cup\Ddef$}
\STATE Update $\vtheta$ \wrt objective in (\ref{eq::1}).
\ENDFOR
\ENDFOR
\STATE $\triangleright$ \quad Inference
\FOR {each input sample $\vx$}
\STATE Predict its label by $h^{-1}(f_{\vtheta}(\vx\oplus\Delta_1))$.
\ENDFOR

\end{algorithmic}
\end{algorithm}

%% file: section/experiment.tex
\subsection{Experiment setting}

\paragraph{Backdoor attack.}
To assess our method, we consider seven leading backdoor attacks: BadNets \citep{gu2019badnets}, Blended method \citep{chen2017targeted}, Sinusoidal Signal (SIG) attacks \citep{barni2019new}, Sample-Specific Backdoor Attacks (SSBA) \citep{li2021invisible}, WaNet \citep{nguyen2021wanet}, BPP attack \citep{Wang_2022_CVPR} and TrojanNN attack \citep{Trojannn}. Note that to expand our evaluation scope, we have modified certain attacks originally intended for training-controllable scenarios by excluding their training control components and we postpone the details to  \app~\ref{app:exp_detail}. For a consistent and reliable evaluation, we utilize configurations from the BackdoorBench framework \citep{wubackdoorbench, wu2024backdoorbench}, which offers a standardized backdoor attack assessment platform. Each attack is implemented with a 5\% poisoning rate, targeting the $0^{th}$ label if not specified. The performance of these attacks is measured across three benchmark dataset, \ie, CIFAR-10 \citep{krizhevsky2009learning}, Tiny ImageNet \citep{le2015tiny}, and GTSRB \citep{stallkamp2011german}, and analyzed using three neural network architectures, \ie, PreAct-ResNet18 \citep{he2016identity} VGG19-BN \citep{simonyan2014very} and ViT-B-16 \citep{dosovitskiy2020image}. Due to limitations in space, we present results for GTSRB and VGG19-BN in \app~\ref{app::add}. It is important to note that the clean label attack SIG is only applicable to CIFAR-10 with the set poisoning ratio. Additional information on these attacks is available in \app~\ref{app:exp_detail}.

\paragraph{Backdoor defense.}
In this paper, we benchmark our approach against popular and advanced backdoor defense methods, including AC \citep{chen2019detecting}, Spectral signatures \citep{tran2018spectral}, ABL \citep{li2021anti}, DBD \citep{huang2022backdoor}, NAB \citep{liu2023beating}. For a fair comparison, we adopt the configurations recommended by the BackdoorBench framework \citep{wubackdoorbench, wu2024backdoorbench}. Note that we were unable to achieve satisfactory results for DBD on Tiny ImageNet with ViT-B-16, so it has been excluded in this case. For our method, we set the reserved dataset size to 10\% of the training dataset unless otherwise specified. The chosen parameters are $\lambda_1 = 1$ and $\lambda_2 = 1$. To enhance the defensive backdoor, each defensive poisoned sample is sampled five times in an epoch, and we set $\tau(\vx) = \vx + 0.1 \cdot \veps$ with $\veps \sim \mathcal{N}(0,1)$. The defensive backdoor utilizes a target mapping function $h(y) = (y + 1) \mod K$, along with a patch trigger with pixel value $2$ as illustrated in Figure~\ref{pi}. More details on the defense methods and supplementary experiments are postponed in \app~\ref{app:exp_detail} and \app~\ref{app::add}.

\paragraph{Metrics.}
To measure the effectiveness of each defense method, we employ three key metrics: Accuracy on benign data ($\textbf{ACC}$), Attack Success Rate ($\textbf{ASR}$), and Defense Effectiveness Rating ($\textbf{DER}$). ACC is the metric indicating model's performance for predicting the benign samples correctly, while ASR quantifies the proportion of poisoned samples that are incorrectly classified to the attacker's intended target label. For our method, the ACC is measured by predicting the benign samples with a defensive trigger to the defensive target, or equivalently reversing the prediction of the benign sample with a defensive trigger to the true label. A higher ACC and a lower ASR signify successful backdoor mitigation.

The DER, used in \citep{Zhu_2023_ICCV, wei2024shared}, is a metric ranging from 0 to 1, designed to evaluate the trade-off between maintaining ACC and reducing ASR. It is defined by the following equation:
\begin{equation}
    \text{DER} = [\max(0, \Delta \text{ASR}) - \max(0,\Delta \text{ACC}) + 1] / 2,
\end{equation}
where $\Delta \text{ASR}$ and $\Delta \text{ACC}$ represent the respective decreases in ASR and ACC between model without defense and model with defense.

\textbf{Note}: Superior defense methods are characterized by higher $\textbf{ACC}$, lower $\textbf{ASR}$, and higher $\textbf{DER}$. In the forthcoming experimental results, the best and second-best performing methods are denoted with \textbf{boldface} and \underline{underline}, respectively.

\subsection{Main results}

\input{table/cifar10_preactresnet18_0_05}

\input{table/tiny_vit_b_16_0_05}

Table~\ref{cifar10_preactresnet18_05} and Table~\ref{tiny_vit_b_16_05} show the proposed method's defense performance compared with other methods, from which we can find:

\textbf{PDB achieves consistent efficacy in mitigating backdoor threats across various attacks, datasets and models.} Specifically, PDB achieves the top-2 lowest ASR across five out of seven attacks on the CIFAR-10 dataset. It also ranks top-2 across all attacks on the GTSRB (Table~\ref{gtsrb_preactresnet18_05}) and Tiny ImageNet. This consistent performance underscores PDB's ability to generalize well across different datasets and attacks. For AC and Spectral, both methods rely on the latent representation of images to detect poisoned samples. AC identifies poisoned samples through clustering in the latent space, considering smaller clusters as likely to contain poisoned data. Spectral detects outliers in the latent space to identify such samples. However, with a poisoning ratio of 5\% for Tiny ImageNet (200 classes, each class accounts for 0.5\%), the poisoned samples become the majority within the target class, breaking the underlying assumptions of both methods and resulting in high ASR values. Additionally, while ABL, DBD, and NAB can defend against certain attacks, they fall short against more sophisticated adversaries, highlighting PDB's robust defense performance.

\textbf{PDB achieves an excellent balance between defense performance and model utility.}
Apart from its robust defensive performance, PDB distinguishes itself through its ability to preserve benign accuracy. Unlike ABL, DBD, and NAB, which often sacrifice considerable benign accuracy in exchange for reduced ASR, leading to lower DER values, PDB maintains a high DER by effectively managing this trade-off. The preservation of model utility, without compromising defense effectiveness, further solidifies PDB's status as a promising strategy in backdoor defense.

The results demonstrate the superiority of PDB in defending against backdoor attacks. By effectively reducing ASR and maintaining a high DER, PDB stands out as a valuable defense approach for backdoor attack.

\subsection{Analysis}

\paragraph{Understanding the effect of PDB.} To elucidate the underlying mechanism of PDB, we delve into the impact of the defensive backdoor by analyzing the T-SNE embeddings and the Trigger Activation Change (TAC). TAC, adapted from \citet{zheng2022data}, is designed to measure the change of activation values for each neuron when comparing maliciously poisoned samples to their benign counterparts. Let $\phi$ be a feature extractor which maps an input image $x$ to the latent activations. For an input image $x$, we can construct the malicious poisoned sample $x\oplus \Delta$. In PDB, a defensive trigger is added to the malicious poisoned sample, crafting sample $x\oplus\Delta\oplus\Delta_1$, aiming to suppress the malicious backdoor. Therefore, for dataset $D$, we define

\begin{align}
&\text{TAC w/o } \Delta_1 = \frac{\sum_{x\in D}(\phi(x\oplus\Delta)-\phi(x))}{|D|}, \\
&\text{TAC w/ } \Delta_1 = \frac{\sum_{x\in D}(\phi(x\oplus\Delta\oplus\Delta_1)-\phi(x))}{|D|}.    
\end{align}

\begin{figure}[H]
    \centering
    \includegraphics[width=\linewidth]{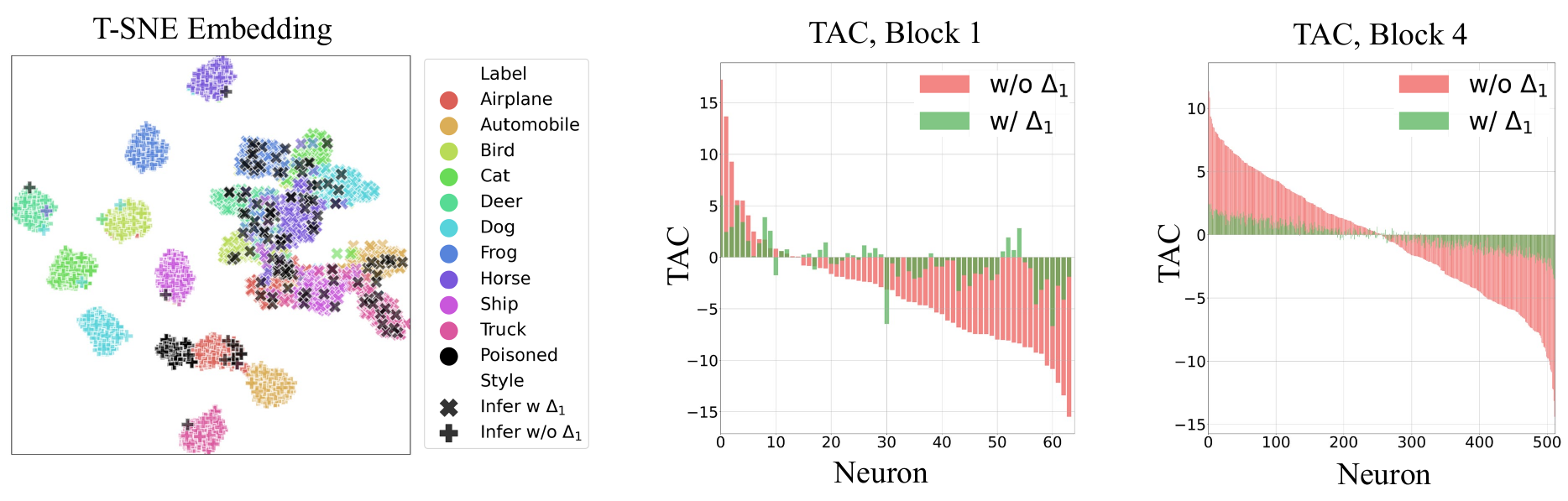}

    \caption{Visualization of T-SNE and TAC  for the BadNets attack on CIFAR-10 with a poisoning ratio of 5\% and PreAct-ResNet. The T-SNE visualizes features in the 4th block of PreAct-ResNet18, and TAC is calculated for both the 1st and the 4th blocks (4 blocks in total). Neurons are indexed in descending order based on their TAC values without $\Delta_1$.}

    \label{fig:under}
\end{figure}

In Figure~\ref{fig:under}, we present the visualization results for the BadNets attack on the CIFAR-10 dataset, utilizing a poisoning ratio of 5\% alongside a PreAct-ResNet architecture. The illustration reveals that planting a defensive trigger to the inputs prompts a shift in the feature space, resulting in the formation of new clusters and effectively alleviating the backdoor effect. Moreover, the TAC analysis for both the initial and final blocks demonstrates that the incorporation of a defensive trigger substantially mitigates the activation changes triggered by the malicious backdoor.

\paragraph{Defense effectiveness under different poisoning ratios.}
To investigate the influence of poisoning ratios on defense performance, we evaluate our method against attacks with poisoning ratios ranging from 1\% to 40\% on CIFAR-10 with PreAct-ResNet18. The results are summarized in Table~\ref{ratio}, from which we can find that the proposed method can consistently mitigate malicious backdoor effect across a wide range of poisoning ratios. For a more comprehensive evaluation of the influence of the poisoning ratio, please refer to \app~\ref{app::add}.

\input{table/ratio}

\paragraph{Training cost comparison.} We first analyze the training complexity of PDB and we refer readers to BackdoorBench\cite{wubackdoorbench} for the training complexity of other methods. Let $C_{sl}$ be the supervised training complexity. Then, we denote the size of the training dataset and the size of the defensive poisoned dataset by $N_{tr}$ and $N_{def}$, respectively. Let $F$ be the frequency of sampling defensive poisoned samples. The training complexity of PDB is given by:
 $O\left(\left(1+\frac{F \cdot N_{def}}{N_{tr}}\right) \cdot C_{sl}\right)$.

To evaluate the empirical runtime, \ie, training time of different defense methods, we conduct experiments  against the BadNets attack for the PreAct-ResNet18 architecture on CIFAR-10 and GTSRB, ViT-B-16 for Tiny ImageNet, all with a poisoning ratio of 5\%. The experiments are conducted on an RTX 4090Ti GPU, and the results are summarized in Table~\ref{time}. From Table~\ref{time}, We can find since $\frac{F \cdot N_{def}}{N_{tr}}$ is set as a small value, the runtime of PDB is not much larger than the baseline (\ie, No Defense). 
In contrast, the runtime of DBD and NAB are significantly higher due to their reliance on self-supervised and semi-supervised training techniques.

\begin{table}[h]

\centering
\caption{Running time (s) comparison of defense methods.}
\label{time}
\setlength{\tabcolsep}{3pt} 
\renewcommand{\arraystretch}{1.5} 
\scalebox{0.62}{
\begin{tabular}{@{}c|c|c|c|c|c|c|c@{}}
\toprule
Defense $\rightarrow$ & No Defense & AC \cite{chen2019detecting}   & Spectral \cite{tran2018spectral} & ABL \cite{li2021anti}  & DBD \cite{huang2022backdoor}   & NAB \cite{liu2023beating}   & PDB (\textbf{Ours})   \\ \midrule
GTSRB    & 801        & 637  & 1642     & 1634 & 7495  & 3081  & 1573  \\
CIFAR-10 & 919        & 658  & 2278     & 1805 & 8351  & 3679  & 1853 \\
Tiny     & 2938       & 2524 & 9952     & 6034 & 26932 & 11698 & 4913  \\ \bottomrule
\end{tabular}}

\end{table}

\paragraph{Resistance to ALL2ALL attack.}  We also evaluate PDB for ALL2ALL attacks on CIFAR-10 using PreAct-ResNet18. The poisoning ratio is set to 5\% and the target labels for samples with labels $y$ are $(y+2)\mod K$ (different from the defensive target). The experimental results are summarized in Table~\ref{a2a_s}. Notably, PDB achieves the best defending performance, demonstrating superior effectiveness in defending against backdoor attacks with multiple targets.

\input{table/a2a_small}

\paragraph{Resistance to adaptive attack.} In our previous experiments, we assumed that attackers had no knowledge of the defense method. However, when attackers are aware of the deployment of PDB, they may design adaptive attacks to bypass the defense. One straightforward approach is to strengthen the malicious backdoor to counteract the defensive backdoor. To assess our method’s resistance to such adaptive attacks, we evaluate it against BadNets with varying trigger sizes and poisoning ratios, representing different strengths of backdoor attacks. The results, summarized in Table~\ref{reb_apt}, demonstrate that PDB can consistently mitigate backdoor against adaptive attacks with various malicious trigger size and poisoning ratios. Note that to keep the stealthness of malicious backdoor, its poisoning ratio and trigger size is expected to be constrained. However, the defensive backdoor can utilize a large trigger size and high sampling frequency to meet the Principle 4, therefore, mitigating the malicious backdoor.

\input{table/reb_apt}

\paragraph{Appendix structure.}
Due to page limitations, more experiments and analyses have been moved to the Appendix. The Appendix is structured as follows: In \app~\ref{app:exp_detail}, we provide the details for the experiments, including the implementation of our method, the parameters, and the setting for all attacks and defense methods. In \app~\ref{app::add}, we provide a more comprehensive comparison between our method and baselines across different datasets, poisoning ratios, and model structures. In \app~\ref{app:ana}, we discuss the influence of key components for PDB, such as triggers, targets, and reserved datasets, and make comparisons to more baselines.

%% file: table/cifar10_preactresnet18_0_05.tex
\begin{table}[H]
\centering
\caption{Results (\%) on CIFAR-10 with PreAct-ResNet18 and poisoning ratio $5.0\%$.}
\label{cifar10_preactresnet18_05}
\setlength{\tabcolsep}{3pt} 
\renewcommand{\arraystretch}{1.5} 
\scalebox{0.62}{
\begin{tabular}{c|cc|ccc|ccc|ccc|ccc|ccc|ccc}
\toprule
Defense $\rightarrow$ & \multicolumn{2}{c|}{No Defense} & \multicolumn{3}{c|}{AC \cite{chen2019detecting}} & \multicolumn{3}{c|}{Spectral \cite{tran2018spectral}} & \multicolumn{3}{c|}{ABL \cite{li2021anti}} & \multicolumn{3}{c|}{DBD \cite{huang2022backdoor}} & \multicolumn{3}{c|}{NAB \cite{liu2023beating}} & \multicolumn{3}{c}{PDB (\textbf{Ours})} \\ \midrule
Attack $\downarrow$ & \multicolumn{1}{c}{ACC} & \multicolumn{1}{c|}{ASR} & \multicolumn{1}{c}{ACC} & \multicolumn{1}{c}{ASR} & \multicolumn{1}{c|}{DER} & \multicolumn{1}{c}{ACC} & \multicolumn{1}{c}{ASR} & \multicolumn{1}{c|}{DER} & \multicolumn{1}{c}{ACC} & \multicolumn{1}{c}{ASR} & \multicolumn{1}{c|}{DER} & \multicolumn{1}{c}{ACC} & \multicolumn{1}{c}{ASR} & \multicolumn{1}{c|}{DER} & \multicolumn{1}{c}{ACC} & \multicolumn{1}{c}{ASR} & \multicolumn{1}{c|}{DER} & \multicolumn{1}{c}{ACC} & \multicolumn{1}{c}{ASR} & \multicolumn{1}{c}{DER} \\
\midrule
BadNets \cite{gu2019badnets} & $92.64$& $88.74$& $90.27$& $75.61$& $55.38$& $\textbf{91.21}$& $69.4$& $58.96$& $84.08$& $\textbf{0.00}$& $90.09$& $87.68$& $2.11$& $\underline{90.84}$& $79.36$& $\underline{0.33}$& $87.57$& $\underline{91.08}$& $0.38$& $\textbf{93.40}$\\
Blended \cite{chen2017targeted} & $93.67$& $99.61$& $\textbf{91.54}$& $98.93$& $49.27$& $\underline{91.43}$& $99.42$& $48.97$& $65.78$& $\textbf{0.00}$& $85.86$& $75.0$& $99.99$& $40.66$& $90.21$& $\underline{0.34}$& $\underline{97.90}$& $91.36$& $0.70$& $\textbf{98.30}$\\
SIG \cite{barni2019new} & $93.64$& $97.09$& $90.24$& $93.32$& $50.18$& $\underline{91.79}$& $96.36$& $49.44$& $46.14$& $\underline{0.00}$& $74.79$& $74.86$& $95.58$& $41.37$& $90.71$& $\textbf{0.00}$& $\underline{97.08}$& $\textbf{91.79}$& $0.06$& $\textbf{97.59}$\\
SSBA \cite{li2021invisible} & $93.27$& $94.91$& $88.47$& $92.64$& $48.73$& $\textbf{92.01}$& $93.19$& $50.23$& $81.6$& $\textbf{0.09}$& $91.58$& $72.19$& $11.34$& $81.24$& $90.52$& $1.07$& $\underline{95.55}$& $\underline{91.58}$& $\underline{0.46}$& $\textbf{96.38}$\\
WaNet \cite{nguyen2021wanet} & $91.76$& $85.5$& $\textbf{91.96}$& $88.72$& $50.0$& $91.47$& $83.84$& $50.68$& $69.49$& $95.20$& $38.86$& $72.22$& $9.93$& $78.01$& $85.17$& $\underline{2.16}$& $\underline{88.38}$& $\underline{91.47}$& $\textbf{0.92}$& $\textbf{92.14}$\\
BPP \cite{Wang_2022_CVPR} & $91.47$& $99.34$& $89.64$& $97.96$& $49.78$& $\textbf{92.10}$& $99.82$& $50.0$& $82.89$& $99.93$& $45.71$& $81.71$& $99.98$& $45.12$& $82.86$& $\underline{76.94}$& $\underline{56.90}$& $\underline{90.43}$& $\textbf{1.90}$& $\textbf{98.20}$\\
Trojan \cite{Trojannn} & $93.79$& $99.99$& $\underline{89.40}$& $99.93$& $47.83$& $86.30$& $99.38$& $46.56$& $18.64$& $100.00$& $12.42$& $72.34$& $100.0$& $39.27$& $87.41$& $\underline{1.16}$& $\underline{96.23}$& $\textbf{91.78}$& $\textbf{0.58}$& $\textbf{98.70}$\\
Average & $92.89$& $95.03$& $90.22$& $92.45$& $50.17$& $\underline{90.90}$& $91.63$& $50.69$& $64.09$& $42.17$& $62.76$& $76.57$& $59.85$& $59.50$& $86.61$& $\underline{11.71}$& $\underline{88.51}$& $\textbf{91.36}$& $\textbf{0.71}$& $\textbf{96.39}$\\

\bottomrule

\end{tabular}
}
\end{table}

%% file: table/tiny_vit_b_16_0_05.tex
\begin{table}[H]
\centering
\caption{Results (\%) on Tiny ImageNet with ViT-B-16 and poisoning ratio $5.0\%$.}
\label{tiny_vit_b_16_05}
\setlength{\tabcolsep}{3pt} 
\renewcommand{\arraystretch}{1.5} 
\scalebox{0.62}{
\begin{tabular}{c|cc|ccc|ccc|ccc|ccc|ccc}
\toprule
Defense $\rightarrow$ & \multicolumn{2}{c|}{No Defense} & \multicolumn{3}{c|}{AC \cite{chen2019detecting}} & \multicolumn{3}{c|}{Spectral \cite{tran2018spectral}} & \multicolumn{3}{c|}{ABL \cite{li2021anti}} & \multicolumn{3}{c|}{NAB \cite{liu2023beating}} & \multicolumn{3}{c}{PDB (\textbf{Ours})} \\ \midrule
Attack $\downarrow$ & \multicolumn{1}{c}{ACC} & \multicolumn{1}{c|}{ASR} & \multicolumn{1}{c}{ACC} & \multicolumn{1}{c}{ASR} & \multicolumn{1}{c|}{DER} & \multicolumn{1}{c}{ACC} & \multicolumn{1}{c}{ASR} & \multicolumn{1}{c|}{DER} & \multicolumn{1}{c}{ACC} & \multicolumn{1}{c}{ASR} & \multicolumn{1}{c|}{DER} & \multicolumn{1}{c}{ACC} & \multicolumn{1}{c}{ASR} & \multicolumn{1}{c|}{DER} & \multicolumn{1}{c}{ACC} & \multicolumn{1}{c}{ASR} & \multicolumn{1}{c}{DER} \\
\midrule
BadNets \cite{gu2019badnets} & $76.15$& $99.72$& $\underline{75.66}$& $99.49$& $49.87$& $74.18$& $\underline{99.47}$& $49.14$& $\textbf{78.19}$& $99.79$& $\underline{50.00}$& $28.88$& $99.83$& $26.36$& $73.71$& $\textbf{0.00}$& $\textbf{98.64}$\\
Blended \cite{chen2017targeted} & $76.00$& $99.83$& $\underline{77.58}$& $99.77$& $50.03$& $76.30$& $\underline{99.62}$& $\underline{50.11}$& $\textbf{78.17}$& $99.93$& $50.00$& $38.43$& $99.79$& $31.24$& $72.70$& $\textbf{0.00}$& $\textbf{98.26}$\\
SSBA \cite{li2021invisible} & $75.30$& $98.86$& $76.01$& $\underline{97.01}$& $\underline{50.93}$& $\underline{76.96}$& $98.73$& $50.07$& $\textbf{78.40}$& $99.59$& $50.00$& $42.03$& $99.37$& $33.36$& $72.65$& $\textbf{0.00}$& $\textbf{98.11}$\\
WaNet \cite{nguyen2021wanet} & $60.90$& $99.74$& $\underline{75.68}$& $\underline{92.46}$& $\underline{53.64}$& $74.27$& $95.91$& $51.91$& $\textbf{77.62}$& $95.37$& $52.18$& $21.89$& $99.76$& $30.50$& $72.82$& $\textbf{0.01}$& $\textbf{99.86}$\\
BPP \cite{Wang_2022_CVPR} & $63.08$& $99.69$& $\underline{76.89}$& $\underline{95.23}$& $\underline{52.23}$& $76.58$& $95.92$& $51.88$& $\textbf{78.19}$& $96.55$& $51.57$& $30.37$& $96.76$& $35.11$& $73.30$& $\textbf{0.00}$& $\textbf{99.84}$\\
Trojan \cite{Trojannn} & $74.98$& $99.77$& $\underline{77.94}$& $\underline{99.78}$& $50.00$& $75.36$& $99.84$& $50.00$& $\textbf{78.40}$& $99.92$& $\underline{50.00}$& $24.15$& $100.00$& $24.58$& $73.00$& $\textbf{0.00}$& $\textbf{98.89}$\\
Average & $71.07$& $99.60$& $\underline{76.63}$& $\underline{97.29}$& $\underline{51.12}$& $75.61$& $98.25$& $50.52$& $\textbf{78.16}$& $98.52$& $50.63$& $30.96$& $99.25$& $30.19$& $73.03$& $\textbf{0.00}$& $\textbf{98.94}$\\

\bottomrule

\end{tabular}
}

\end{table}

%% file: table/ratio.tex
\begin{table}[H]
\centering
\caption{Defense results (\%) under different poisoning ratios on CIFAR-10 and PreAct-ResNet18.}
\label{ratio}
\setlength{\tabcolsep}{3pt} 
\renewcommand{\arraystretch}{1.5} 
\scalebox{0.62}{
\begin{tabular}{@{}c|cc|cc|cc|cc|cc|cc|cc|cc|cc|cc@{}}
\toprule
Poisoning ratio $\rightarrow$ & \multicolumn{4}{c|}{1\%}                                 & \multicolumn{4}{c|}{5\%}                                 & \multicolumn{4}{c|}{10\%}                                 & \multicolumn{4}{c|}{20\%}                                 & \multicolumn{4}{c}{40\%}                                 \\
\midrule
Defense $\rightarrow$        & \multicolumn{2}{c|}{No Defense} & \multicolumn{2}{c|}{PDB (\textbf{Ours})} & \multicolumn{2}{c|}{No Defense} & \multicolumn{2}{c|}{PDB (\textbf{Ours})} & \multicolumn{2}{c|}{No Defense} & \multicolumn{2}{c|}{PDB (\textbf{Ours})} & \multicolumn{2}{c|}{No Defense} & \multicolumn{2}{c|}{PDB (\textbf{Ours})} & \multicolumn{2}{c|}{No Defense} & \multicolumn{2}{c}{PDB (\textbf{Ours})} \\
\midrule
Attack $\downarrow$          & ACC            & ASR           & ACC         & ASR       & ACC            & ASR           & ACC         & ASR       & ACC            & ASR           & ACC         & ASR       & ACC            & ASR           & ACC         & ASR       & ACC            & ASR           & ACC          & ASR      \\
\midrule
BadNets \cite{gu2019badnets}          & 93.14          & 74.73         & 91.59       & 0.31      & 92.64          & 88.74         & 91.08       & 0.38      & 91.32          & 95.03         & 90.25       & 0.40      & 90.17          & 96.12         & 89.18       & 0.16      & 86.16          & 97.77         & 86.32       & 0.28      \\
Blended \cite{chen2017targeted}         & 93.76          & 94.88         & 91.77       & 1.20      & 93.67          & 99.61         & 91.36       & 0.70      & 93.47          & 99.92         & 91.21       & 0.92      & 92.92          & 99.92         & 90.74       & 1.51      & 91.74          & 99.98         & 89.04       & 0.27      \\
BPP \cite{Wang_2022_CVPR}             & 90.81          & 87.23         & 90.73       & 1.38      & 91.47          & 99.34         & 90.43       & 1.90      & 90.69          & 99.78         & 90.47       & 1.11      & 91.45          & 99.71         & 90.44       & 1.29      & 90.66          & 99.99         & 89.22       & 0.49      \\
Average         & 92.57          & 85.61         & 91.36       & 0.96      & 92.59          & 95.90         & 90.96       & 0.99      & 91.83          & 98.24         & 90.64       & 0.81      & 91.51          & 98.58         & 90.12       & 0.99      & 89.52          & 99.25         & 88.19       & 0.34      \\ \bottomrule
\end{tabular}}
\end{table}

%% file: table/a2a_small.tex
\begin{table}[H]
\centering
\caption{ALL2ALL attack results (\%) on CIFAR-10 with PreAct-ResNet18 and poisoning ratio $5.0\%$.}
\label{a2a_s}
\setlength{\tabcolsep}{3pt} 
\renewcommand{\arraystretch}{1.5} 
\scalebox{0.62}{
\begin{tabular}{c|cc|ccc|ccc|ccc|ccc|ccc|ccc}
\toprule
Defense $\rightarrow$ & \multicolumn{2}{c|}{No Defense} & \multicolumn{3}{c|}{AC \cite{chen2019detecting}} & \multicolumn{3}{c|}{Spectral \cite{tran2018spectral}} & \multicolumn{3}{c|}{ABL \cite{li2021anti}} & \multicolumn{3}{c|}{DBD \cite{huang2022backdoor}} & \multicolumn{3}{c|}{NAB \cite{liu2023beating}} & \multicolumn{3}{c}{PDB (\textbf{Ours})} \\ \midrule
Attack $\downarrow$ & \multicolumn{1}{c}{ACC} & \multicolumn{1}{c|}{ASR} & \multicolumn{1}{c}{ACC} & \multicolumn{1}{c}{ASR} & \multicolumn{1}{c|}{DER} & \multicolumn{1}{c}{ACC} & \multicolumn{1}{c}{ASR} & \multicolumn{1}{c|}{DER} & \multicolumn{1}{c}{ACC} & \multicolumn{1}{c}{ASR} & \multicolumn{1}{c|}{DER} & \multicolumn{1}{c}{ACC} & \multicolumn{1}{c}{ASR} & \multicolumn{1}{c|}{DER} & \multicolumn{1}{c}{ACC} & \multicolumn{1}{c}{ASR} & \multicolumn{1}{c|}{DER} & \multicolumn{1}{c}{ACC} & \multicolumn{1}{c}{ASR} & \multicolumn{1}{c}{DER} \\
\midrule
BadNets \cite{gu2019badnets} & $92.50$& $61.33$& $90.10$& $53.7$& $52.61$& $\textbf{92.33}$& $57.73$& $51.72$& $52.46$& $59.96$& $30.66$& $87.10$& $\underline{4.52}$& $\underline{75.70}$& $80.51$& $62.74$& $44.00$& $\underline{90.68}$& $\textbf{2.72}$& $\textbf{78.40}$\\
Blended \cite{chen2017targeted} & $93.51$& $83.87$& $91.36$& $78.56$& $51.58$& $\textbf{93.72}$& $84.66$& $50.00$& $68.04$& $35.62$& $61.39$& $75.24$& $\underline{26.62}$& $\underline{69.49}$& $90.34$& $79.09$& $50.80$& $\underline{91.87}$& $\textbf{3.95}$& $\textbf{89.14}$\\
SIG \cite{barni2019new} & $93.52$& $88.15$& $91.49$& $83.07$& $51.52$& $\textbf{94.02}$& $88.77$& $50.00$& $67.20$& $59.67$& $51.08$& $76.19$& $\underline{20.26}$& $\underline{75.28}$& $82.65$& $83.19$& $47.04$& $\underline{91.73}$& $\textbf{3.13}$& $\textbf{91.62}$\\

\bottomrule

\end{tabular}}
\end{table}

%% file: table/reb_apt.tex
\begin{table}[!ht]
    \centering
    \caption{Defense results (\%) against adaptive attacks with different poisoning ratios.}
    \label{reb_apt}
    \setlength{\tabcolsep}{3pt} 
\renewcommand{\arraystretch}{1.5} 
\scalebox{0.62}{
    \begin{tabular}{c|cc|cc|cc|cc|cc|cc}
        \toprule
        Poisoning ratio → & \multicolumn{4}{c|}{10\%} & \multicolumn{4}{c|}{20\%} & \multicolumn{4}{c}{30\%} \\ \midrule
        Defense → & \multicolumn{2}{c|}{No Defense} & \multicolumn{2}{c|}{PDB (\textbf{Ours})} &  \multicolumn{2}{c|}{No Defense} & \multicolumn{2}{c|}{PDB (\textbf{Ours})} &  \multicolumn{2}{c|}{No Defense} & \multicolumn{2}{c}{PDB (\textbf{Ours})} \\ \midrule
        Malicious trigger size ↓ & ACC & ASR & ACC & ASR & ACC & ASR & ACC & ASR & ACC & ASR & ACC & ASR \\ \midrule
        4x4 & 92.39 & 96.83 & 90.66 & 0.18 & 91.14 & 97.67 & 90.01 & 0.21 & 90.38 & 98.13 & 89.65 & 0.49 \\ 
        5x5 & 93.11 & 97.69 & 91.28 & 0.29 & 92.79 & 97.98 & 90.97 & 0.28 & 92.20 & 98.30 & 90.02 & 0.56 \\ 
        6x6 & 93.26 & 98.16 & 91.62 & 0.27 & 92.48 & 98.68 & 90.83 & 0.33 & 92.01 & 98.83 & 90.03 & 0.69 \\ 
        7x7 & 93.65 & 98.66 & 91.46 & 0.31 & 93.07 & 99.03 & 91.03 & 0.56 & 92.56 & 99.23 & 90.48 & 0.67 \\ 
        8x8 & 93.51 & 99.24 & 91.16 & 0.37 & 92.82 & 99.38 & 91.14 & 0.58 & 92.53 & 99.50 & 90.27 & 0.74 \\ 
        9x9 & 93.45 & 99.53 & 91.12 & 0.51 & 92.76 & 99.67 & 90.84 & 0.56 & 92.15 & 99.72 & 90.39 & 0.67 \\ 
        10x10 & 93.20 & 99.66 & 91.37 & 0.54 & 93.17 & 99.74 & 90.76 & 0.78 & 92.58 & 99.81 & 90.45 & 0.82 \\ \bottomrule
    \end{tabular}}
\end{table}

%% file: section/conclusion.tex
In this paper, we propose a proactive approach to defend against malicious backdoor attacks in DNNs. Rather than relying on traditional detection and mitigation pipeline, our method, PDB, leverages the \textit{“home field”} advantage of defenders to inject a defensive backdoor to fight against malicious backdoor. To achieve such a goal, we introduce four essential design properties for an effective defensive backdoor: reversibility, inaccessibility to attackers, minimal impact on model performance, and resistance to other backdoors. By incorporating a defensive backdoor during training, we suppress the impact of malicious backdoors when the defensive trigger is present. Our approach offers several advantages over existing methods. First, it does not rely on accurate detection of poisoned samples and any assumption for attacks, avoiding performance degradation when some poisoned samples evade detection. Second, PDB does not require complex modifications to the training process, minimizing training cost. In summary, PDB provides a novel and effective defense method against backdoor attacks, enhancing the safety and reliability of DNNs. 
\vspace{-0.05in}

\paragraph{Limitations and future work.} Currently, PDB faces several key limitations. First, its reliance on clean samples presents a practical challenge, prompting the exploration of alternative sources, such as generated data. Second, investigating PDB across diverse machine learning tasks is essential for broader applicability. Addressing these limitations through future research will enhance the defense’s effectiveness and facilitate its widespread adoption in safeguarding machine learning systems against backdoor attacks.
\vspace{-0.05in}

\paragraph{Broader impacts.} 
The broader impacts can be considered from both positive and negative perspectives. On the positive side, PDB enhances the security and reliability of DNNs, thereby contributing to the trustworthiness of AI technologies. However, there are potential negative implications that should be considered. The technique could potentially be misused if it falls into the wrong hands, who might use the defensive backdoor mechanism for nefarious purposes.

%% file: appendix/exp_detail.tex
\section{Experiment details}
\label{app:exp_detail}
In our experiments, we adapted all baselines and settings from \href{https://github.com/SCLBD/BackdoorBench}{BackdoorBench} \citep{wubackdoorbench}. Moreover, all checkpoints of attack methods are sourced from BackdoorBench and the defense results are aligned with the leaderboard in BackdoorBench if applicable. Below, we outline the details of various backdoor attacks:

\subsection{Attack details}
\begin{itemize}
    \item BadNets \citep{gu2019badnets} is one of the earliest works for backdoor learning, which inserts a small patch of fixed pattern to replace some pixels in the image. We use the default setting in BackdoorBench. 

    \item Blended backdoor attack (Blended) \citep{chen2017targeted} uses an alpha-blending strategy to fuse images with fixed patterns. We set $\alpha=0.2$ as the default in BackdoorBench. Note that a large $\alpha$ causes visual-perceptible changes to clean samples, making the Blended Attack challenging for defense methods.
   
    \item Sinusoidal signal backdoor attack (SIG) \citep{barni2019new} is a clean-label attack that perturbs clean images in the target label using a sinusoidal signal as the trigger.  We use the default setting in BackdoorBench.
    
    \item Sample-specific backdoor attack (SSBA) \citep{li2021invisible} uses an auto-encoder to fuse a trigger into clean samples and generate poisoned samples. We use the default setting in BackdoorBench.
    
    \item Warping-based poisoned networks (WaNet) \citep{nguyen2021wanet} is also a training-controllable attack that perturbs clean samples using a warping function to construct poisoned samples. We use the default setting in BackdoorBench.
    
   \item Bppattack (BPP) \citep{Wang_2022_CVPR} is also a training-controllable attack that employs image quantization and dithering as the Trojan trigger.  We use the default setting in BackdoorBench.

   \item Trojaning attack on neural networks (TrojanNN) \citep{Trojannn} inverses the neural network to generate a general trojan trigger. We use the default setting in BackdoorBench.
\end{itemize}

\paragraph{Adaptation to data poisoning attack.} In our paper, we explore scenarios where attacks can only utilize data poisoning techniques. To facilitate a more comprehensive comparison of our method, we modify attacks originally designed for training-controllable scenarios, removing the training component to adapt them to a data poisoning setting.

\subsection{Defense details}
Here, we summarize the details of each defense method used:
\begin{itemize}    
    \item  AC \citep{chen2019detecting} is a detection method that detects the poisoned sample using the abnormal clustering for poisoned samples. By removing the detected samples, AC can effectively defend against backdoor attack. We use the default setting in BackdoorBench.

   \item  Spectral \citep{tran2018spectral} is a detection method that detects the poisoned sample using the abnormal Spectral Signature for poisoned samples. By removing the detected samples, AC can effectively defend against backdoor attack. We use the default setting in BackdoorBench.

   \item  ABL \citep{li2021anti} utilizes the early-learning phenomenon of poisoned samples to detect poisoned samples and then unlearns them to mitigate the backdoor effect. We use the default setting in BackdoorBench.

   \item  DBD \citep{huang2022backdoor} divides the training process into three stages and uses self-supervised techniques to detect the poisoned sample and learn a clean model. We use the default setting in BackdoorBench.

   \item  NAB \citep{liu2023beating} first employs an advanced detection method to filter the poisoned samples. Then, the detected samples are relabeled by employing other techniques and planted with non-adversarial triggers to suppress the backdoor. In this work, we use the detection method from ABL and the self-supervised method from DBD to relabel the samples. For other settings, We use the default setting in BackdoorBench.
    
    \item  FT finetunes the model on a small, clean, reserved dataset to mitigate the backdoor effect. We use the default setting in BackdoorBench.
    
    \item  FP \citep{liu2018fine} is a pruning-based method that prunes neurons according to their activations and then fine-tunes the model to keep clean accuracy. We use the default setting in BackdoorBench.

    \item  NC \citep{wang2019neural} first optimizes a possible trigger to detect backdoored models. If detected as backdoored, unlearn the optimized trigger. If detected as clean, the model is returned unchanged.
    
    \item NAD \citep{li2021neural} uses Attention Distillation to mitigate backdoors.  We use the default setting in BackdoorBench.

    \item i-BAU \citep{zeng2022adversarial} uses adversarial training with UAP and hyper-gradient to mitigate the backdoor.  We use the default setting in BackdoorBench. 

    \item PDB (\textbf{Ours}) defends backdoor attack by injecting defensive backdoor. We set the reserved dataset size to 10\% of the training dataset. The chosen parameters are $\lambda_1 = 1$ and $\lambda_2 = 1$. To enhance the defensive backdoor, each defensive poisoned sample is sampled five times in an epoch, and we set $\tau(\vx) = \vx + 0.1 \cdot \veps$ with $\veps \sim \mathcal{N}(0,1)$. The defensive backdoor utilizes a target mapping function $h(y) = (y + 1) \mod K$, along with a $7 \times 7$ patch trigger with pixel value $2$ as illustrated in Figure~\ref{pi}. 
 
\end{itemize}

\paragraph{Adaptation to ViT-B-16.} For all experiments on CIFAR-10 and GTSRB, we train the model 100 epochs with batch size 256 for fair comparison. For Tiny ImageNet with ViT-B-16, we consider a fine-tuning task as recommended by BackdoorBench. Specifically, we train each model 10 epochs with batch size 128 and initialize the model with pre-trained weights.

%% file: appendix/exp_add.tex
\section{Additional experiment results}
\label{app::add}

This section provides additional experiment results to supplement the observations claimed in Section~\ref{main_exp}.

\subsection{Main experiments on GTSRB with PreAct-ResNet18}
Table \ref{gtsrb_preactresnet18_05}, \ref{gtsrb_preactresnet18_1}, and \ref{gtsrb_preactresnet18_01} summarize the results of various defense methods against backdoor attacks on the GTSRB dataset using the PreAct-ResNet18 model architecture. These methods were evaluated at different poisoning ratios (1.0\%, 5.0\%, and 10.0\%). Notably, the results demonstrate that PDB effectively mitigates backdoor attacks, consistently achieving top-2 defense performance across all cases.
\input{table/gtsrb_preactresnet18_0_05}
\input{table/gtsrb_preactresnet18_0_1}
\input{table/gtsrb_preactresnet18_0_01}

\subsection{Main experiments on CIFAR-10 with PreAct-ResNet18}
Table \ref{cifar10_preactresnet18_1} and \ref{cifar10_preactresnet18_01} summarize the results of different defense methods against backdoor attacks on the CIFAR-10 dataset using the PreAct-ResNet18 model architecture. These methods were evaluated at different poisoning ratios (1.0\% and 10.0\%). Notably, they achieved the top-2 lowest ASR in 12 out of 14 cases.

\input{table/cifar10_preactresnet18_0_1}
\input{table/cifar10_preactresnet18_0_01}

\subsection{Main experiments on CIFAR-10 with VGG19-BN}
Table \ref{cifar10_vgg19_bn_05}, \ref{cifar10_vgg19_bn_01}, and \ref{cifar10_vgg19_bn_1} summarize the results of different defense methods against backdoor attacks on the CIFAR-10 dataset using the VGG19-BN model architecture. These methods were evaluated at different poisoning ratios (1.0\%, 5\%, and 10.0\%). Impressively, PDB achieves the top-2 lowest ASR in 20 out of 21 cases.

\input{table/cifar10_vgg19_bn_0_05}
\input{table/cifar10_vgg19_bn_0_01}
\input{table/cifar10_vgg19_bn_0_1}

\subsection{Main experiments on Tiny ImageNet with ViT-B-16}
To demonstrate the effectiveness and scalability of PDB, we evaluate our proposed method against backdoor attacks on the Tiny ImageNet dataset using the ViT-B-6 model architecture. We consider different poisoning ratios (1.0\%, 5\%, and 10.0\%). The results, presented in Table~\ref{vit_all}, highlight our method’s ability to effectively mitigate backdoor attacks for a large dataset and a large model.

\input{table/vit_all}

\subsection{Experiments on invisible backdoor attack and low-poisoning ratio attack}
As aforementioned, the proposed method, PDB, does not rely on specific assumptions about the type of attack, making it effective for defending against both invisible backdoor attacks and attacks with low poisoning ratios. To demonstrate this effectiveness, we conducted experiments using low poisoning ratios (0.5\% and 0.1\%) for both \textit{Visible} and \textit{Invisible} attacks. The results are summarized in Table~\ref{reb_inv}, from which we can find that PDB can consistently mitigate backdoor attacks.

\input{table/reb_inv}

%% file: table/gtsrb_preactresnet18_0_05.tex
\begin{table}[H]
\centering
\caption{Results on GTSRB with PreAct-ResNet18 and poisoning ratio $5.0\%$.}
\label{gtsrb_preactresnet18_05}
\setlength{\tabcolsep}{3pt} 
\renewcommand{\arraystretch}{1.5} 
\scalebox{0.62}{
\begin{tabular}{c|cc|ccc|ccc|ccc|ccc|ccc|ccc}
\toprule
Defense $\rightarrow$ & \multicolumn{2}{c|}{No Defense} & \multicolumn{3}{c|}{AC \cite{chen2019detecting}} & \multicolumn{3}{c|}{Spectral \cite{tran2018spectral}} & \multicolumn{3}{c|}{ABL \cite{li2021anti}} & \multicolumn{3}{c|}{DBD \cite{huang2022backdoor}} & \multicolumn{3}{c|}{NAB \cite{liu2023beating}} & \multicolumn{3}{c}{PDB (\textbf{Ours})} \\ \midrule
Attack $\downarrow$ & \multicolumn{1}{c}{ACC} & \multicolumn{1}{c|}{ASR} & \multicolumn{1}{c}{ACC} & \multicolumn{1}{c}{ASR} & \multicolumn{1}{c|}{DER} & \multicolumn{1}{c}{ACC} & \multicolumn{1}{c}{ASR} & \multicolumn{1}{c|}{DER} & \multicolumn{1}{c}{ACC} & \multicolumn{1}{c}{ASR} & \multicolumn{1}{c|}{DER} & \multicolumn{1}{c}{ACC} & \multicolumn{1}{c}{ASR} & \multicolumn{1}{c|}{DER} & \multicolumn{1}{c}{ACC} & \multicolumn{1}{c}{ASR} & \multicolumn{1}{c|}{DER} & \multicolumn{1}{c}{ACC} & \multicolumn{1}{c}{ASR} & \multicolumn{1}{c}{DER} \\
\midrule
BadNets \cite{gu2019badnets} & $97.3$& $57.95$& $\textbf{97.55}$& $84.67$& $50.0$& $96.36$& $92.51$& $49.53$& $95.8$& $0.00$& $\underline{78.22}$& $78.76$& $\underline{0.00}$& $69.70$& $84.06$& $92.32$& $43.38$& $\underline{96.75}$& $\textbf{0.00}$& $\textbf{78.70}$\\
Blended \cite{chen2017targeted} & $98.84$& $99.93$& $\underline{97.48}$& $99.73$& $49.42$& $\textbf{97.58}$& $99.83$& $49.41$& $44.18$& $\textbf{0.00}$& $\underline{72.63}$& $83.08$& $100.00$& $42.12$& $87.94$& $99.34$& $44.84$& $97.00$& $\underline{0.03}$& $\textbf{99.03}$\\
SSBA \cite{li2021invisible} & $98.11$& $99.34$& $97.16$& $98.3$& $50.05$& $\underline{97.31}$& $97.96$& $50.29$& $80.2$& $\textbf{0.01}$& $90.71$& $81.57$& $99.96$& $41.73$& $94.14$& $13.57$& $\underline{90.90}$& $\textbf{97.61}$& $\underline{0.02}$& $\textbf{99.41}$\\
WaNet \cite{nguyen2021wanet} & $97.42$& $92.85$& $95.91$& $54.1$& $68.62$& $\underline{96.85}$& $62.12$& $65.08$& $1.45$& $99.40$& $2.02$& $84.09$& $\underline{0.01}$& $\underline{89.76}$& $84.96$& $9.53$& $85.43$& $\textbf{96.92}$& $\textbf{0.00}$& $\textbf{96.17}$\\
BPP \cite{Wang_2022_CVPR} & $98.21$& $98.97$& $\textbf{97.64}$& $83.67$& $57.37$& $97.36$& $87.3$& $55.42$& $12.77$& $100.00$& $7.28$& $82.52$& $0.00$& $91.64$& $89.39$& $\underline{0.00}$& $\underline{95.08}$& $\underline{97.46}$& $\textbf{0.00}$& $\textbf{99.11}$\\
Trojan \cite{Trojannn} & $98.55$& $100.0$& $97.00$& $100.0$& $49.22$& $\textbf{97.94}$& $100.0$& $49.7$& $83.42$& $\underline{0.00}$& $\underline{92.43}$& $73.17$& $0.01$& $87.30$& $89.32$& $12.10$& $89.33$& $\underline{97.73}$& $\textbf{0.00}$& $\textbf{99.59}$\\
Average & $98.07$& $91.51$& $97.12$& $86.74$& $54.11$& $\underline{97.23}$& $89.95$& $53.24$& $52.97$& $\underline{33.23}$& $57.22$& $80.53$& $33.33$& $70.38$& $88.3$& $37.81$& $\underline{74.83}$& $\textbf{97.24}$& $\textbf{0.01}$& $\textbf{95.33}$\\

\bottomrule

\end{tabular}
}
\end{table}

%% file: table/gtsrb_preactresnet18_0_1.tex
\begin{table}[H]
\centering
\caption{Results on GTSRB with PreAct-ResNet18 and poisoning ratio $10.0\%$.}
\label{gtsrb_preactresnet18_1}
\setlength{\tabcolsep}{3pt} 
\renewcommand{\arraystretch}{1.5} 
\scalebox{0.62}{
\begin{tabular}{c|cc|ccc|ccc|ccc|ccc|ccc|ccc}
\toprule
Defense $\rightarrow$ & \multicolumn{2}{c|}{No Defense} & \multicolumn{3}{c|}{AC \cite{chen2019detecting}} & \multicolumn{3}{c|}{Spectral \cite{tran2018spectral}} & \multicolumn{3}{c|}{ABL \cite{li2021anti}} & \multicolumn{3}{c|}{DBD \cite{huang2022backdoor}} & \multicolumn{3}{c|}{NAB \cite{liu2023beating}} & \multicolumn{3}{c}{PDB (\textbf{Ours})} \\ \midrule
Attack $\downarrow$ & \multicolumn{1}{c}{ACC} & \multicolumn{1}{c|}{ASR} & \multicolumn{1}{c}{ACC} & \multicolumn{1}{c}{ASR} & \multicolumn{1}{c|}{DER} & \multicolumn{1}{c}{ACC} & \multicolumn{1}{c}{ASR} & \multicolumn{1}{c|}{DER} & \multicolumn{1}{c}{ACC} & \multicolumn{1}{c}{ASR} & \multicolumn{1}{c|}{DER} & \multicolumn{1}{c}{ACC} & \multicolumn{1}{c}{ASR} & \multicolumn{1}{c|}{DER} & \multicolumn{1}{c}{ACC} & \multicolumn{1}{c}{ASR} & \multicolumn{1}{c|}{DER} & \multicolumn{1}{c}{ACC} & \multicolumn{1}{c}{ASR} & \multicolumn{1}{c}{DER} \\
\midrule
BadNets \cite{gu2019badnets} & $97.24$& $59.25$& $95.53$& $95.94$& $49.14$& $\underline{96.83}$& $94.22$& $49.79$& $93.82$& $0.00$& $77.92$& $82.75$& $\underline{0.00}$& $72.38$& $95.34$& $0.03$& $\underline{78.66}$& $\textbf{97.05}$& $\textbf{0.00}$& $\textbf{79.53}$\\
Blended \cite{chen2017targeted} & $98.58$& $99.99$& $\underline{98.61}$& $100.0$& $50.0$& $\textbf{98.76}$& $100.0$& $50.0$& $32.99$& $\textbf{0.00}$& $67.2$& $81.73$& $99.97$& $41.58$& $95.91$& $2.87$& $\underline{97.23}$& $96.98$& $\underline{0.02}$& $\textbf{99.18}$\\
SSBA \cite{li2021invisible} & $97.98$& $99.56$& $96.60$& $99.47$& $49.36$& $\textbf{97.57}$& $99.55$& $49.8$& $63.21$& $\underline{0.58}$& $82.1$& $91.11$& $99.94$& $46.56$& $\underline{96.82}$& $0.85$& $\underline{98.77}$& $96.42$& $\textbf{0.00}$& $\textbf{99.00}$\\
WaNet \cite{nguyen2021wanet} & $97.74$& $94.25$& $96.36$& $72.65$& $60.11$& $\underline{96.94}$& $74.82$& $59.32$& $21.35$& $84.09$& $16.88$& $84.03$& $\underline{0.00}$& $\underline{90.27}$& $82.88$& $75.26$& $52.07$& $\textbf{97.36}$& $\textbf{0.00}$& $\textbf{96.94}$\\
BPP \cite{Wang_2022_CVPR} & $97.43$& $99.9$& $\underline{97.43}$& $88.69$& $55.6$& $\textbf{97.73}$& $93.32$& $53.29$& $10.51$& $99.94$& $6.54$& $86.47$& $100.00$& $44.52$& $81.92$& $\underline{32.01}$& $\underline{76.19}$& $97.40$& $\textbf{0.00}$& $\textbf{99.93}$\\
Trojan \cite{Trojannn} & $98.57$& $100.0$& $96.76$& $100.0$& $49.1$& $\underline{97.73}$& $100.0$& $49.58$& $78.19$& $\underline{0.00}$& $89.81$& $87.09$& $100.00$& $44.26$& $\textbf{97.75}$& $0.06$& $\textbf{99.56}$& $96.90$& $\textbf{0.00}$& $\underline{99.17}$\\
Average & $97.92$& $92.16$& $96.88$& $92.79$& $52.22$& $\textbf{97.59}$& $93.65$& $51.96$& $50.01$& $30.77$& $56.74$& $85.53$& $66.65$& $56.60$& $91.77$& $\underline{18.51}$& $\underline{83.75}$& $\underline{97.02}$& $\textbf{0.00}$& $\textbf{95.63}$\\

\bottomrule

\end{tabular}
}
\end{table}

%% file: table/gtsrb_preactresnet18_0_01.tex
\begin{table}[H]
\centering
\caption{Results on GTSRB with PreAct-ResNet18 and poisoning ratio $1.0\%$.}
\label{gtsrb_preactresnet18_01}
\setlength{\tabcolsep}{3pt} 
\renewcommand{\arraystretch}{1.5} 
\scalebox{0.62}{
\begin{tabular}{c|cc|ccc|ccc|ccc|ccc|ccc|ccc}
\toprule
Defense $\rightarrow$ & \multicolumn{2}{c|}{No Defense} & \multicolumn{3}{c|}{AC \cite{chen2019detecting}} & \multicolumn{3}{c|}{Spectral \cite{tran2018spectral}} & \multicolumn{3}{c|}{ABL \cite{li2021anti}} & \multicolumn{3}{c|}{DBD \cite{huang2022backdoor}} & \multicolumn{3}{c|}{NAB \cite{liu2023beating}} & \multicolumn{3}{c}{PDB (\textbf{Ours})} \\ \midrule
Attack $\downarrow$ & \multicolumn{1}{c}{ACC} & \multicolumn{1}{c|}{ASR} & \multicolumn{1}{c}{ACC} & \multicolumn{1}{c}{ASR} & \multicolumn{1}{c|}{DER} & \multicolumn{1}{c}{ACC} & \multicolumn{1}{c}{ASR} & \multicolumn{1}{c|}{DER} & \multicolumn{1}{c}{ACC} & \multicolumn{1}{c}{ASR} & \multicolumn{1}{c|}{DER} & \multicolumn{1}{c}{ACC} & \multicolumn{1}{c}{ASR} & \multicolumn{1}{c|}{DER} & \multicolumn{1}{c}{ACC} & \multicolumn{1}{c}{ASR} & \multicolumn{1}{c|}{DER} & \multicolumn{1}{c}{ACC} & \multicolumn{1}{c}{ASR} & \multicolumn{1}{c}{DER} \\
\midrule
BadNets \cite{gu2019badnets} & $98.35$& $50.26$& $\textbf{98.42}$& $81.73$& $50.00$& $\underline{97.32}$& $86.26$& $49.49$& $0.48$& $100.00$& $1.06$& $84.29$& $\underline{0.00}$& $\underline{68.10}$& $87.16$& $3.51$& $67.78$& $96.62$& $\textbf{0.00}$& $\textbf{74.27}$\\
Blended \cite{chen2017targeted} & $98.8$& $95.67$& $97.10$& $94.34$& $49.82$& $\underline{97.70}$& $94.02$& $50.28$& $28.22$& $\textbf{2.27}$& $\underline{61.41}$& $83.4$& $99.97$& $42.30$& $86.6$& $97.14$& $43.9$& $\textbf{98.20}$& $\underline{6.13}$& $\textbf{94.47}$\\
SSBA \cite{li2021invisible} & $98.75$& $94.54$& $96.83$& $86.5$& $53.06$& $\underline{97.28}$& $87.65$& $52.71$& $4.12$& $68.09$& $15.91$& $88.6$& $\textbf{0.00}$& $\underline{92.20}$& $88.08$& $95.41$& $44.67$& $\textbf{98.06}$& $\underline{0.49}$& $\textbf{96.68}$\\
WaNet \cite{nguyen2021wanet} & $97.08$& $62.24$& $96.48$& $4.14$& $\underline{78.75}$& $\textbf{97.35}$& $28.02$& $67.11$& $29.75$& $33.21$& $30.86$& $87.15$& $\underline{0.00}$& $76.16$& $85.29$& $2.7$& $73.87$& $\underline{97.30}$& $\textbf{0.00}$& $\textbf{81.12}$\\
BPP \cite{Wang_2022_CVPR} & $98.26$& $62.21$& $\underline{97.11}$& $40.91$& $60.08$& $\textbf{98.12}$& $64.61$& $49.93$& $6.41$& $20.48$& $24.94$& $78.41$& $\underline{0.00}$& $\underline{71.18}$& $87.55$& $55.82$& $47.85$& $96.52$& $\textbf{0.00}$& $\textbf{80.24}$\\
Trojan \cite{Trojannn} & $98.17$& $100.0$& $\underline{97.38}$& $100.0$& $49.60$& $\textbf{97.72}$& $100.0$& $49.77$& $21.69$& $0.00$& $61.76$& $81.74$& $\underline{0.00}$& $\underline{91.78}$& $84.39$& $99.82$& $43.2$& $96.83$& $\textbf{0.00}$& $\textbf{99.33}$\\
Average & $98.23$& $77.49$& $97.22$& $67.94$& $56.89$& $\textbf{97.58}$& $76.76$& $53.22$& $15.11$& $37.34$& $32.66$& $83.93$& $\underline{16.66}$& $\underline{73.62}$& $86.51$& $59.07$& $53.54$& $\underline{97.26}$& $\textbf{1.10}$& $\textbf{87.69}$\\

\bottomrule

\end{tabular}
}
\end{table}

%% file: table/cifar10_preactresnet18_0_1.tex
\begin{table}[H]
\centering
\caption{Results on CIFAR-10 with PreAct-ResNet18 and poisoning ratio $10.0\%$.}
\label{cifar10_preactresnet18_1}
\setlength{\tabcolsep}{3pt} 
\renewcommand{\arraystretch}{1.5} 
\scalebox{0.62}{
\begin{tabular}{c|cc|ccc|ccc|ccc|ccc|ccc|ccc}
\toprule
Defense $\rightarrow$ & \multicolumn{2}{c|}{No Defense} & \multicolumn{3}{c|}{AC \cite{chen2019detecting}} & \multicolumn{3}{c|}{Spectral \cite{tran2018spectral}} & \multicolumn{3}{c|}{ABL \cite{li2021anti}} & \multicolumn{3}{c|}{DBD \cite{huang2022backdoor}} & \multicolumn{3}{c|}{NAB \cite{liu2023beating}} & \multicolumn{3}{c}{PDB (\textbf{Ours})} \\ \midrule
Attack $\downarrow$ & \multicolumn{1}{c}{ACC} & \multicolumn{1}{c|}{ASR} & \multicolumn{1}{c}{ACC} & \multicolumn{1}{c}{ASR} & \multicolumn{1}{c|}{DER} & \multicolumn{1}{c}{ACC} & \multicolumn{1}{c}{ASR} & \multicolumn{1}{c|}{DER} & \multicolumn{1}{c}{ACC} & \multicolumn{1}{c}{ASR} & \multicolumn{1}{c|}{DER} & \multicolumn{1}{c}{ACC} & \multicolumn{1}{c}{ASR} & \multicolumn{1}{c|}{DER} & \multicolumn{1}{c}{ACC} & \multicolumn{1}{c}{ASR} & \multicolumn{1}{c|}{DER} & \multicolumn{1}{c}{ACC} & \multicolumn{1}{c}{ASR} & \multicolumn{1}{c}{DER} \\
\midrule
BadNets \cite{gu2019badnets} & $91.32$& $95.03$& $88.80$& $86.23$& $53.14$& $\underline{89.98}$& $92.41$& $50.64$& $83.32$& $\textbf{0.00}$& $93.52$& $89.65$& $1.28$& $\underline{96.04}$& $75.51$& $\underline{0.08}$& $89.57$& $\textbf{90.25}$& $0.40$& $\textbf{96.78}$\\
Blended \cite{chen2017targeted} & $93.47$& $99.92$& $88.52$& $99.72$& $47.63$& $\underline{90.35}$& $99.84$& $48.48$& $77.3$& $\underline{0.73}$& $91.51$& $69.91$& $99.98$& $38.22$& $86.25$& $\textbf{0.12}$& $\underline{96.29}$& $\textbf{91.21}$& $0.92$& $\textbf{98.37}$\\
SIG \cite{barni2019new} & $84.48$& $98.27$& $82.41$& $94.61$& $50.79$& $83.01$& $92.27$& $52.26$& $57.8$& $\underline{0.00}$& $85.79$& $60.67$& $100.0$& $38.10$& $\underline{83.07}$& $17.54$& $\underline{89.66}$& $\textbf{91.10}$& $\textbf{0.00}$& $\textbf{99.13}$\\
SSBA \cite{li2021invisible} & $92.88$& $97.86$& $\underline{90.00}$& $96.23$& $49.37$& $89.63$& $90.5$& $52.05$& $80.79$& $\textbf{0.00}$& $92.88$& $63.5$& $99.51$& $35.31$& $88.77$& $1.84$& $\underline{95.95}$& $\textbf{90.95}$& $\underline{0.19}$& $\textbf{97.87}$\\
WaNet \cite{nguyen2021wanet} & $91.25$& $89.73$& $\underline{91.93}$& $96.8$& $50.0$& $\textbf{91.94}$& $90.17$& $50.0$& $83.19$& $\textbf{0.00}$& $\underline{90.84}$& $80.9$& $6.61$& $86.39$& $80.01$& $0.88$& $88.81$& $90.92$& $\underline{0.29}$& $\textbf{94.56}$\\
BPP \cite{Wang_2022_CVPR} & $90.69$& $99.78$& $89.29$& $99.73$& $49.32$& $\textbf{92.34}$& $99.72$& $50.03$& $78.55$& $13.77$& $86.94$& $68.65$& $100.0$& $38.98$& $75.66$& $\textbf{0.21}$& $\underline{92.27}$& $\underline{90.47}$& $\underline{1.11}$& $\textbf{99.22}$\\
Trojan \cite{Trojannn} & $93.42$& $100.0$& $\underline{89.75}$& $99.97$& $48.18$& $89.70$& $100.0$& $48.14$& $11.07$& $100.00$& $8.82$& $66.23$& $100.0$& $36.40$& $86.17$& $\underline{1.49}$& $\underline{95.63}$& $\textbf{91.24}$& $\textbf{0.59}$& $\textbf{98.62}$\\
Average & $91.07$& $97.23$& $88.67$& $96.19$& $49.78$& $\underline{89.56}$& $94.99$& $50.23$& $67.43$& $16.36$& $78.61$& $71.36$& $72.48$& $52.78$& $82.21$& $\underline{3.17}$& $\underline{92.60}$& $\textbf{90.88}$& $\textbf{0.50}$& $\textbf{97.79}$\\

\bottomrule

\end{tabular}
}
\end{table}

%% file: table/cifar10_preactresnet18_0_01.tex
\begin{table}[H]
\centering
\caption{Results on CIFAR-10 with PreAct-ResNet18 and poisoning ratio $1.0\%$.}
\label{cifar10_preactresnet18_01}
\setlength{\tabcolsep}{3pt} 
\renewcommand{\arraystretch}{1.5} 
\scalebox{0.62}{
\begin{tabular}{c|cc|ccc|ccc|ccc|ccc|ccc|ccc}
\toprule
Defense $\rightarrow$ & \multicolumn{2}{c|}{No Defense} & \multicolumn{3}{c|}{AC \cite{chen2019detecting}} & \multicolumn{3}{c|}{Spectral \cite{tran2018spectral}} & \multicolumn{3}{c|}{ABL \cite{li2021anti}} & \multicolumn{3}{c|}{DBD \cite{huang2022backdoor}} & \multicolumn{3}{c|}{NAB \cite{liu2023beating}} & \multicolumn{3}{c}{PDB (\textbf{Ours})} \\ \midrule
Attack $\downarrow$ & \multicolumn{1}{c}{ACC} & \multicolumn{1}{c|}{ASR} & \multicolumn{1}{c}{ACC} & \multicolumn{1}{c}{ASR} & \multicolumn{1}{c|}{DER} & \multicolumn{1}{c}{ACC} & \multicolumn{1}{c}{ASR} & \multicolumn{1}{c|}{DER} & \multicolumn{1}{c}{ACC} & \multicolumn{1}{c}{ASR} & \multicolumn{1}{c|}{DER} & \multicolumn{1}{c}{ACC} & \multicolumn{1}{c}{ASR} & \multicolumn{1}{c|}{DER} & \multicolumn{1}{c}{ACC} & \multicolumn{1}{c}{ASR} & \multicolumn{1}{c|}{DER} & \multicolumn{1}{c}{ACC} & \multicolumn{1}{c}{ASR} & \multicolumn{1}{c}{DER} \\
\midrule
BadNets \cite{gu2019badnets} & $93.14$& $74.73$& $88.88$& $27.18$& $71.65$& $\textbf{92.62}$& $74.2$& $50.01$& $72.81$& $53.20$& $50.6$& $78.09$& $2.99$& $78.35$& $90.85$& $\underline{1.43}$& $\underline{85.50}$& $\underline{91.59}$& $\textbf{0.31}$& $\textbf{86.44}$\\
Blended \cite{chen2017targeted} & $93.76$& $94.88$& $89.77$& $86.09$& $52.40$& $\textbf{93.27}$& $93.32$& $50.53$& $66.26$& $\textbf{0.17}$& $83.61$& $70.18$& $8.04$& $81.63$& $86.65$& $3.23$& $\underline{92.27}$& $\underline{91.77}$& $\underline{1.20}$& $\textbf{95.84}$\\
SIG \cite{barni2019new} & $93.82$& $83.4$& $90.0$& $74.81$& $52.38$& $\textbf{93.09}$& $85.91$& $49.63$& $64.33$& $\textbf{0.00}$& $76.96$& $75.01$& $67.82$& $48.38$& $88.73$& $2.40$& $\underline{87.96}$& $\underline{90.21}$& $\underline{1.19}$& $\textbf{89.30}$\\
SSBA \cite{li2021invisible} & $93.43$& $73.44$& $89.61$& $27.92$& $70.85$& $\textbf{92.71}$& $54.79$& $58.97$& $83.11$& $56.33$& $53.4$& $78.52$& $\underline{1.13}$& $\underline{78.70}$& $85.11$& $64.51$& $50.31$& $\underline{91.56}$& $\textbf{1.02}$& $\textbf{85.28}$\\
WaNet \cite{nguyen2021wanet} & $90.65$& $12.63$& $89.53$& $4.87$& $\underline{53.32}$& $\textbf{92.79}$& $8.81$& $51.91$& $65.43$& $53.43$& $37.39$& $79.74$& $\underline{4.60}$& $48.56$& $85.91$& $21.92$& $47.63$& $\underline{91.45}$& $\textbf{0.82}$& $\textbf{55.91}$\\
BPP \cite{Wang_2022_CVPR} & $90.81$& $87.23$& $89.13$& $16.74$& $84.40$& $\textbf{91.94}$& $21.02$& $83.11$& $55.61$& $11.78$& $70.13$& $86.3$& $\underline{6.91}$& $\underline{87.91}$& $\underline{91.06}$& $45.62$& $70.81$& $90.73$& $\textbf{1.38}$& $\textbf{92.89}$\\
Trojan \cite{Trojannn} & $93.58$& $99.97$& $89.72$& $99.53$& $48.29$& $\textbf{92.83}$& $99.61$& $49.8$& $24.73$& $100.00$& $15.58$& $74.49$& $99.99$& $40.45$& $88.52$& $\underline{5.67}$& $\underline{94.62}$& $\underline{90.01}$& $\textbf{1.90}$& $\textbf{97.25}$\\
Average & $92.74$& $75.18$& $89.52$& $48.16$& $61.90$& $\textbf{92.75}$& $62.52$& $56.28$& $61.75$& $39.27$& $55.38$& $77.48$& $27.36$& $66.28$& $88.12$& $\underline{20.68}$& $\underline{75.58}$& $\underline{91.05}$& $\textbf{1.12}$& $\textbf{86.13}$\\

\bottomrule

\end{tabular}
}
\end{table}

%% file: table/cifar10_vgg19_bn_0_05.tex
\begin{table}[H]
\centering
\caption{Results on CIFAR-10 with VGG19-BN and poisoning ratio $5.0\%$.}
\label{cifar10_vgg19_bn_05}
\setlength{\tabcolsep}{3pt} 
\renewcommand{\arraystretch}{1.5} 
\scalebox{0.62}{
\begin{tabular}{c|cc|ccc|ccc|ccc|ccc|ccc|ccc}
\toprule
Defense $\rightarrow$ & \multicolumn{2}{c|}{No Defense} & \multicolumn{3}{c|}{AC \cite{chen2019detecting}} & \multicolumn{3}{c|}{Spectral \cite{tran2018spectral}} & \multicolumn{3}{c|}{ABL \cite{li2021anti}} & \multicolumn{3}{c|}{DBD \cite{huang2022backdoor}} & \multicolumn{3}{c|}{NAB \cite{liu2023beating}} & \multicolumn{3}{c}{PDB (\textbf{Ours})} \\ \midrule
Attack $\downarrow$ & \multicolumn{1}{c}{ACC} & \multicolumn{1}{c|}{ASR} & \multicolumn{1}{c}{ACC} & \multicolumn{1}{c}{ASR} & \multicolumn{1}{c|}{DER} & \multicolumn{1}{c}{ACC} & \multicolumn{1}{c}{ASR} & \multicolumn{1}{c|}{DER} & \multicolumn{1}{c}{ACC} & \multicolumn{1}{c}{ASR} & \multicolumn{1}{c|}{DER} & \multicolumn{1}{c}{ACC} & \multicolumn{1}{c}{ASR} & \multicolumn{1}{c|}{DER} & \multicolumn{1}{c}{ACC} & \multicolumn{1}{c}{ASR} & \multicolumn{1}{c|}{DER} & \multicolumn{1}{c}{ACC} & \multicolumn{1}{c}{ASR} & \multicolumn{1}{c}{DER} \\
\midrule
BadNets \cite{gu2019badnets} & $91.19$& $93.92$& $85.75$& $88.49$& $50.0$& $9.97$& $99.99$& $9.39$& $\textbf{90.35}$& $21.18$& $85.95$& $61.15$& $\underline{2.61}$& $80.63$& $75.28$& $2.79$& $\underline{87.61}$& $\underline{87.43}$& $\textbf{1.17}$& $\textbf{94.50}$\\
Blended \cite{chen2017targeted} & $92.24$& $99.43$& $\textbf{87.18}$& $97.82$& $48.28$& $10.0$& $100.0$& $8.88$& $75.92$& $\textbf{0.12}$& $\underline{91.50}$& $52.72$& $99.99$& $30.24$& $76.36$& $6.52$& $88.52$& $\underline{86.14}$& $\underline{1.09}$& $\textbf{96.12}$\\
SIG \cite{barni2019new} & $91.91$& $97.23$& $\underline{86.30}$& $97.11$& $47.26$& $70.18$& $96.84$& $39.33$& $84.29$& $\underline{0.00}$& $\underline{94.81}$& $54.16$& $99.36$& $31.13$& $76.65$& $0.64$& $90.66$& $\textbf{88.77}$& $\textbf{0.00}$& $\textbf{97.05}$\\
SSBA \cite{li2021invisible} & $91.53$& $90.39$& $\underline{85.93}$& $74.07$& $55.36$& $10.0$& $100.0$& $9.23$& $79.72$& $38.47$& $70.06$& $48.43$& $97.05$& $28.45$& $77.14$& $\underline{12.84}$& $\underline{81.58}$& $\textbf{87.34}$& $\textbf{4.87}$& $\textbf{90.67}$\\
WaNet \cite{nguyen2021wanet} & $87.42$& $94.32$& $\underline{86.03}$& $52.47$& $70.23$& $10.0$& $100.0$& $11.29$& $69.85$& $98.71$& $41.22$& $56.72$& $\underline{19.82}$& $\underline{71.90}$& $62.73$& $55.31$& $57.16$& $\textbf{87.87}$& $\textbf{2.15}$& $\textbf{96.09}$\\
BPP \cite{Wang_2022_CVPR} & $89.3$& $98.3$& $\underline{86.50}$& $72.54$& $61.48$& $10.0$& $100.0$& $10.35$& $39.62$& $\textbf{0.00}$& $74.31$& $59.82$& $17.05$& $\underline{75.89}$& $48.10$& $84.27$& $36.42$& $\textbf{87.96}$& $\underline{0.07}$& $\textbf{98.45}$\\
Trojan \cite{Trojannn} & $92.26$& $99.99$& $86.74$& $99.97$& $47.25$& $10.71$& $97.26$& $10.59$& $76.25$& $\underline{0.00}$& $91.99$& $47.8$& $100.00$& $27.77$& $\textbf{88.00}$& $2.68$& $\underline{96.53}$& $\underline{87.61}$& $\textbf{0.00}$& $\textbf{97.67}$\\
Average & $90.84$& $96.23$& $\underline{86.35}$& $83.21$& $54.26$& $18.69$& $99.16$& $14.15$& $73.71$& $\underline{22.64}$& $\underline{78.55}$& $54.4$& $62.27$& $49.43$& $72.04$& $23.58$& $76.92$& $\textbf{87.59}$& $\textbf{1.33}$& $\textbf{95.79}$\\

\bottomrule

\end{tabular}
}
\end{table}

%% file: table/cifar10_vgg19_bn_0_01.tex
\begin{table}[H]
\centering
\caption{Results on CIFAR-10 with VGG19-BN and poisoning ratio $1.0\%$.}
\label{cifar10_vgg19_bn_01}
\setlength{\tabcolsep}{3pt} 
\renewcommand{\arraystretch}{1.5} 
\scalebox{0.62}{
\begin{tabular}{c|cc|ccc|ccc|ccc|ccc|ccc|ccc}
\toprule
Defense $\rightarrow$ & \multicolumn{2}{c|}{No Defense} & \multicolumn{3}{c|}{AC \cite{chen2019detecting}} & \multicolumn{3}{c|}{Spectral \cite{tran2018spectral}} & \multicolumn{3}{c|}{ABL \cite{li2021anti}} & \multicolumn{3}{c|}{DBD \cite{huang2022backdoor}} & \multicolumn{3}{c|}{NAB \cite{liu2023beating}} & \multicolumn{3}{c}{PDB (\textbf{Ours})} \\ \midrule
Attack $\downarrow$ & \multicolumn{1}{c}{ACC} & \multicolumn{1}{c|}{ASR} & \multicolumn{1}{c}{ACC} & \multicolumn{1}{c}{ASR} & \multicolumn{1}{c|}{DER} & \multicolumn{1}{c}{ACC} & \multicolumn{1}{c}{ASR} & \multicolumn{1}{c|}{DER} & \multicolumn{1}{c}{ACC} & \multicolumn{1}{c}{ASR} & \multicolumn{1}{c|}{DER} & \multicolumn{1}{c}{ACC} & \multicolumn{1}{c}{ASR} & \multicolumn{1}{c|}{DER} & \multicolumn{1}{c}{ACC} & \multicolumn{1}{c}{ASR} & \multicolumn{1}{c|}{DER} & \multicolumn{1}{c}{ACC} & \multicolumn{1}{c}{ASR} & \multicolumn{1}{c}{DER} \\
\midrule
BadNets \cite{gu2019badnets} & $91.51$& $79.87$& $86.58$& $9.01$& $\underline{82.96}$& $\textbf{89.80}$& $55.54$& $61.31$& $79.32$& $88.57$& $43.9$& $54.81$& $11.12$& $66.02$& $76.48$& $\underline{5.72}$& $79.56$& $\underline{88.67}$& $\textbf{1.34}$& $\textbf{87.84}$\\
Blended \cite{chen2017targeted} & $91.83$& $94.19$& $87.17$& $85.10$& $52.21$& $\textbf{90.07}$& $93.07$& $49.68$& $85.11$& $95.13$& $46.64$& $55.09$& $\textbf{0.00}$& $78.73$& $67.57$& $10.40$& $\underline{79.76}$& $\underline{87.46}$& $\underline{4.38}$& $\textbf{92.72}$\\
SIG \cite{barni2019new} & $92.11$& $92.79$& $87.22$& $79.34$& $54.28$& $\textbf{89.10}$& $\underline{1.36}$& $\textbf{94.21}$& $87.37$& $92.86$& $47.63$& $55.92$& $\textbf{0.00}$& $78.3$& $76.45$& $88.38$& $44.38$& $\underline{88.31}$& $8.30$& $\underline{90.34}$\\
SSBA \cite{li2021invisible} & $91.9$& $38.08$& $\underline{86.90}$& $\underline{10.61}$& $\underline{61.23}$& $\textbf{89.82}$& $24.12$& $55.94$& $77.15$& $61.7$& $42.62$& $58.45$& $12.20$& $46.21$& $71.51$& $44.13$& $39.80$& $86.02$& $\textbf{2.24}$& $\textbf{64.98}$\\
WaNet \cite{nguyen2021wanet} & $90.12$& $23.47$& $\underline{86.56}$& $\underline{5.19}$& $\underline{57.36}$& $10.00$& $100.00$& $9.94$& $10.03$& $100.0$& $9.96$& $55.8$& $15.22$& $36.97$& $61.62$& $22.64$& $36.17$& $\textbf{88.63}$& $\textbf{1.52}$& $\textbf{60.23}$\\
BPP \cite{Wang_2022_CVPR} & $89.2$& $47.71$& $\underline{86.43}$& $\underline{4.64}$& $\underline{70.15}$& $10.00$& $100.00$& $10.40$& $84.03$& $11.11$& $65.72$& $56.81$& $7.67$& $53.83$& $75.22$& $26.57$& $53.58$& $\textbf{88.36}$& $\textbf{1.36}$& $\textbf{72.76}$\\
Trojan \cite{Trojannn} & $91.97$& $99.92$& $\underline{87.57}$& $99.68$& $47.92$& $\textbf{90.47}$& $28.72$& $84.85$& $80.36$& $99.98$& $44.2$& $55.68$& $99.87$& $31.88$& $68.77$& $\textbf{6.37}$& $\underline{85.18}$& $87.55$& $\underline{7.70}$& $\textbf{93.90}$\\
Average & $91.23$& $68.0$& $\underline{86.92}$& $41.94$& $\underline{60.87}$& $67.04$& $57.54$& $52.33$& $71.91$& $78.48$& $42.95$& $56.08$& $\underline{20.87}$& $55.99$& $71.09$& $29.17$& $59.78$& $\textbf{87.86}$& $\textbf{3.84}$& $\textbf{80.40}$\\

\bottomrule

\end{tabular}
}
\end{table}

%% file: table/cifar10_vgg19_bn_0_1.tex
\begin{table}[H]
\centering
\caption{Results on CIFAR-10 with VGG19-BN and poisoning ratio $10.0\%$.}
\label{cifar10_vgg19_bn_1}
\setlength{\tabcolsep}{3pt} 
\renewcommand{\arraystretch}{1.5} 
\scalebox{0.62}{
\begin{tabular}{c|cc|ccc|ccc|ccc|ccc|ccc|ccc}
\toprule
Defense $\rightarrow$ & \multicolumn{2}{c|}{No Defense} & \multicolumn{3}{c|}{AC \cite{chen2019detecting}} & \multicolumn{3}{c|}{Spectral \cite{tran2018spectral}} & \multicolumn{3}{c|}{ABL \cite{li2021anti}} & \multicolumn{3}{c|}{DBD \cite{huang2022backdoor}} & \multicolumn{3}{c|}{NAB \cite{liu2023beating}} & \multicolumn{3}{c}{PDB (\textbf{Ours})} \\ \midrule
Attack $\downarrow$ & \multicolumn{1}{c}{ACC} & \multicolumn{1}{c|}{ASR} & \multicolumn{1}{c}{ACC} & \multicolumn{1}{c}{ASR} & \multicolumn{1}{c|}{DER} & \multicolumn{1}{c}{ACC} & \multicolumn{1}{c}{ASR} & \multicolumn{1}{c|}{DER} & \multicolumn{1}{c}{ACC} & \multicolumn{1}{c}{ASR} & \multicolumn{1}{c|}{DER} & \multicolumn{1}{c}{ACC} & \multicolumn{1}{c}{ASR} & \multicolumn{1}{c|}{DER} & \multicolumn{1}{c}{ACC} & \multicolumn{1}{c}{ASR} & \multicolumn{1}{c|}{DER} & \multicolumn{1}{c}{ACC} & \multicolumn{1}{c}{ASR} & \multicolumn{1}{c}{DER} \\
\midrule
BadNets \cite{gu2019badnets} & $90.42$& $94.43$& $84.81$& $93.83$& $47.50$& $10.00$& $100.0$& $9.79$& $\textbf{89.50}$& $12.16$& $\underline{90.68}$& $56.35$& $20.95$& $69.71$& $37.38$& $\underline{1.54}$& $69.92$& $\underline{86.03}$& $\textbf{1.47}$& $\textbf{94.29}$\\
Blended \cite{chen2017targeted} & $91.91$& $99.5$& $\underline{86.21}$& $99.17$& $47.32$& $10.00$& $100.0$& $9.04$& $85.18$& $4.60$& $\underline{94.08}$& $46.97$& $99.97$& $27.53$& $35.24$& $\underline{3.01}$& $69.91$& $\textbf{88.04}$& $\textbf{0.24}$& $\textbf{97.69}$\\
SIG \cite{barni2019new} & $83.48$& $98.87$& $79.00$& $99.8$& $47.76$& $27.90$& $98.6$& $22.34$& $31.68$& $\underline{0.00}$& $73.53$& $43.4$& $99.97$& $29.96$& $\textbf{80.64}$& $6.48$& $\underline{94.77}$& $\underline{79.48}$& $\textbf{0.00}$& $\textbf{97.43}$\\
SSBA \cite{li2021invisible} & $90.85$& $95.11$& $85.54$& $90.61$& $49.60$& $10.00$& $100.0$& $9.57$& $\textbf{86.27}$& $\textbf{0.16}$& $\textbf{95.19}$& $51.21$& $98.81$& $30.18$& $74.21$& $2.64$& $87.91$& $\underline{85.82}$& $\underline{0.70}$& $\underline{94.69}$\\
WaNet \cite{nguyen2021wanet} & $84.58$& $96.49$& $\underline{85.05}$& $80.78$& $57.86$& $10.00$& $100.0$& $12.71$& $68.10$& $99.82$& $41.76$& $56.87$& $36.84$& $65.97$& $68.73$& $\underline{36.18}$& $\underline{72.23}$& $\textbf{86.99}$& $\textbf{2.41}$& $\textbf{97.04}$\\
BPP \cite{Wang_2022_CVPR} & $89.32$& $99.79$& $86.62$& $93.22$& $\underline{51.93}$& $\textbf{90.00}$& $99.16$& $50.32$& $72.09$& $100.00$& $41.38$& $54.07$& $99.96$& $32.38$& $52.92$& $\underline{89.00}$& $37.19$& $\underline{87.79}$& $\textbf{0.12}$& $\textbf{99.07}$\\
Trojan \cite{Trojannn} & $91.57$& $100.0$& $\underline{87.07}$& $99.97$& $47.77$& $10.00$& $100.0$& $9.22$& $81.70$& $\underline{0.00}$& $\underline{95.06}$& $46.6$& $100.0$& $27.51$& $80.45$& $2.21$& $93.33$& $\textbf{87.73}$& $\textbf{0.00}$& $\textbf{98.08}$\\
Average & $88.88$& $97.74$& $\underline{84.90}$& $93.91$& $49.96$& $23.99$& $99.68$& $17.57$& $73.50$& $30.96$& $\underline{75.96}$& $50.78$& $79.5$& $40.46$& $61.37$& $\underline{20.15}$& $75.04$& $\textbf{85.98}$& $\textbf{0.71}$& $\textbf{96.90}$\\

\bottomrule

\end{tabular}
}
\end{table}

%% file: table/vit_all.tex
\begin{table}[H]
\centering
\caption{Results on Tiny ImageNet with ViT-B-16.}
\label{vit_all}
\setlength{\tabcolsep}{10pt} 
\renewcommand{\arraystretch}{1.5} 
\scalebox{0.62}{
\begin{tabular}{c|cc|cc|cc|cc|cc|cc}
\toprule
Poisoning ratio $\rightarrow$ & \multicolumn{4}{c|}{$10\%$}                                 & \multicolumn{4}{c|}{$5\%$}                                 & \multicolumn{4}{c}{$1\%$}                                 \\ \midrule
Defense $\rightarrow$        & \multicolumn{2}{c|}{No Defense} & \multicolumn{2}{c|}{PDB (\textbf{Ours})} & \multicolumn{2}{c|}{No Defense|} & \multicolumn{2}{c|}{PDB (\textbf{Ours})} & \multicolumn{2}{c|}{No Defense} & \multicolumn{2}{c}{PDB (\textbf{Ours})} \\
\midrule
Attack $\downarrow$          & ACC            & ASR           & ACC         & ASR       & ACC            & ASR           & ACC         & ASR       & ACC            & ASR           & ACC         & ASR       \\
\midrule
BadNets \cite{gu2019badnets}          & 73.96          & 99.79         & 72.88       & 0.01      & 76.15          & 99.72         & 73.71       & 0.00      & 76.98          & 97.45         & 74.87       & 0.01      \\
Blended \cite{chen2017targeted}         & 75.28          & 99.93         & 73.85       & 0.00      & 76.00          & 99.83         & 72.70       & 0.00      & 77.39          & 98.03         & 74.24       & 0.00      \\
SSBA \cite{li2021invisible}            & 76.37          & 99.28         & 74.27       & 0.00      & 75.30          & 99.86         & 72.65       & 0.00      & 76.05          & 88.96         & 74.62       & 0.00      \\
WaNet \cite{nguyen2021wanet}           & 62.68          & 99.61         & 74.46       & 0.00      & 60.90          & 99.73         & 72.82       & 0.01      & 63.40          & 95.87         & 74.49       & 0.05      \\
BPP \cite{Wang_2022_CVPR}             & 61.81          & 99.82         & 72.83       & 0.00      & 63.08          & 99.68         & 73.30       & 0.00      & 62.98          & 94.21         & 73.84       & 0.00      \\
Trojan \cite{Trojannn}        & 75.16          & 99.86         & 72.72       & 0.00      & 74.98          & 99.76         & 73.00       & 0.00      & 77.15          & 99.07         & 75.01       & 0.00      \\
Average         & 70.88          & 99.72         & 73.50       & 0.00      & 71.07          & 98.86         & 73.03       & 0.00      & 72.33          & 95.03         & 74.51       & 0.01     \\ \bottomrule
\end{tabular}}
\end{table}

%% file: table/reb_inv.tex
\begin{table}[!ht]
    \centering
    \caption{Results on PreAct-ResNet18 and CIFAR10 for invisible and low poisoning  ratio attacks}
    \label{reb_inv}
        \setlength{\tabcolsep}{3pt} 
\renewcommand{\arraystretch}{1.5} 
\scalebox{0.62}{
    \begin{tabular}{c|c|c|c|c|c|c|c|c|c|c|c}
        \toprule
        ~ & Trigger Type → & Visible & Visible & Invisible & Invisible & Invisible & Invisible & Invisible & Invisible & Invisible & Invisible \\ \midrule
        Attack → & ~ & BadNet & BadNet & Blended & Blended & Sig & Sig & SSBA & SSBA & WaNet & WaNet \\ 
        Poisoning ratio ↓ & Defense ↓ & ACC & ASR & ACC & ASR & ACC & ASR & ACC & ASR & ACC & ASR \\ 
        0.10\% & No Defense & 93.61 & 1.23 & 93.80 & 56.11 & 93.73 & 41.27 & 93.89 & 1.62 & 92.18 & 0.78 \\ 
        0.10\% & PDB & 91.55 & 0.87 & 91.91 & 0.36 & 91.59 & 0.39 & 91.72 & 0.42 & 91.87 & 0.89 \\ 
        0.50\% & No Defense & 93.76 & 50.06 & 93.68 & 93.30 & 93.80 & 82.43 & 93.41 & 35.67 & 91.27 & 1.12 \\ 
        0.50\% & PDB & 91.62 & 0.60 & 91.66 & 0.31 & 91.72 & 0.12 & 91.65 & 0.54 & 91.72 & 0.92 \\ \bottomrule
    \end{tabular}}
\end{table}

%% file: appendix/add_analysis.tex
\section{Discussion and additional analysis}
\label{app:ana}

\subsection{Influences of reserved dataset}
We explore the impact of the reserved dataset on defense performance, considering both dataset size and source:
\begin{itemize}[leftmargin=*]
    \item \textbf{Dataset size.} We investigate the influence of reserved dataset size using the CIFAR-10 dataset, a 5\% poisoning ratio, and the PreAct-ResNet architecture. Figure~\ref{fig:ratio} summarizes the results, demonstrating that our proposed method effectively mitigates malicious backdoor effects across a wide range of reserved dataset sizes. Notably, increasing the reserved dataset significantly improves the accuracy of our method. 
    
    \begin{figure}[H]
    \centering
    \includegraphics[width=\linewidth]{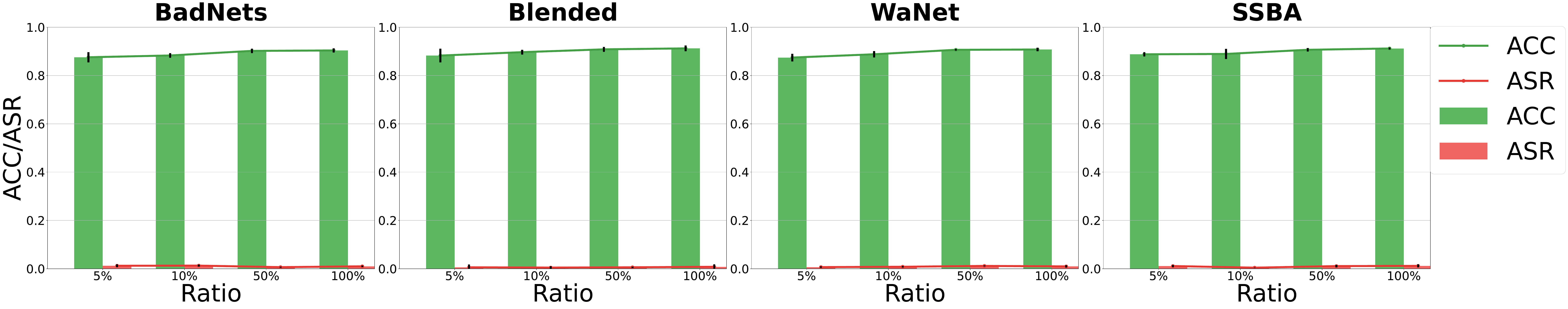}
    \caption{Defense results with different sizes of the reserved dataset. The ratio represents the ratio of the reserved dataset compared with the whole training dataset. Note that the Defense ASR is below 1\% and may be barely visible.}
    \label{fig:ratio}
    \end{figure}

    \item \textbf{Defense with generated dataset.} As generative models have evolved, incorporating the generated dataset as an additional source for backdoor defense has become practical and reasonable. To explore the source of this reserved dataset, we assess our method using the generated data CIFAR-5m  from \citet{nakkiran2021the}. with DDPM. Our findings, summarized in Table~\ref{fake}, demonstrate that the advanced generative model can indeed supply a sufficient dataset for applying our defense strategy.
    \input{table/fake}
\end{itemize}

\subsection{Influences of the trigger and target}
\begin{figure}[H]
    \centering
    \includegraphics[width=0.8\linewidth]{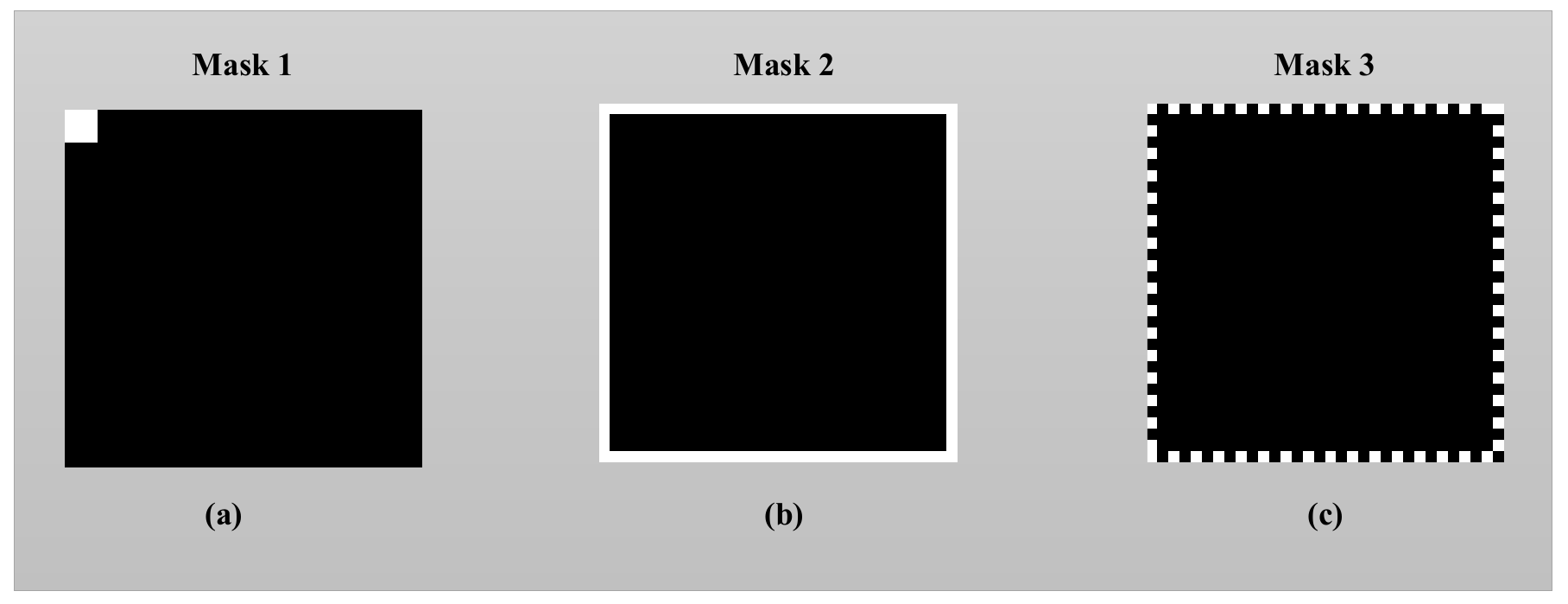}
    \caption{The masks of patched triggers.}
    \label{trigger}
\end{figure}
In this section, we explore the impact of trigger design and target assignment strategy. Specifically, we evaluate different trigger configurations using a patch trigger with three distinct masks (illustrated in Figure~\ref{trigger}). Additionally, we consider two target assignment strategies, \ie, $h_1(y)=(y+1)\mod K$ and $h_2(\phi(\vx)) = -\phi(\vx)$, where $y$ is the hard label and $\phi(\vx)$ is the logits for the model output. Our experiments, detailed in Table~\ref{config}, demonstrate the effectiveness of our method across various defensive backdoor designs.

\input{table/config}

\paragraph{Special case study: Same target label for malicious backdoor and defensive backdoor.} 
While our trigger remains inaccessible to the attacker, the target assignment strategy could potentially be stolen or coincidentally used by the attacker. In this scenario, we conduct experiments where both the attacker and the defender employ the same target assignment strategy but different triggers. Our results demonstrate that our method remains effective in such cases. Specifically, assuming both the attacker and defender choose $h(y) = (y + 1) \mod K$, we summarize the results in Table~\ref{a2a_1}, highlighting our method's resilience even when the attacker uses the same target label as the defender.

\input{table/a2a_1}

\subsection{Influences of augmentation}
\begin{figure}[H]
    \centering
    \includegraphics[width=\linewidth]{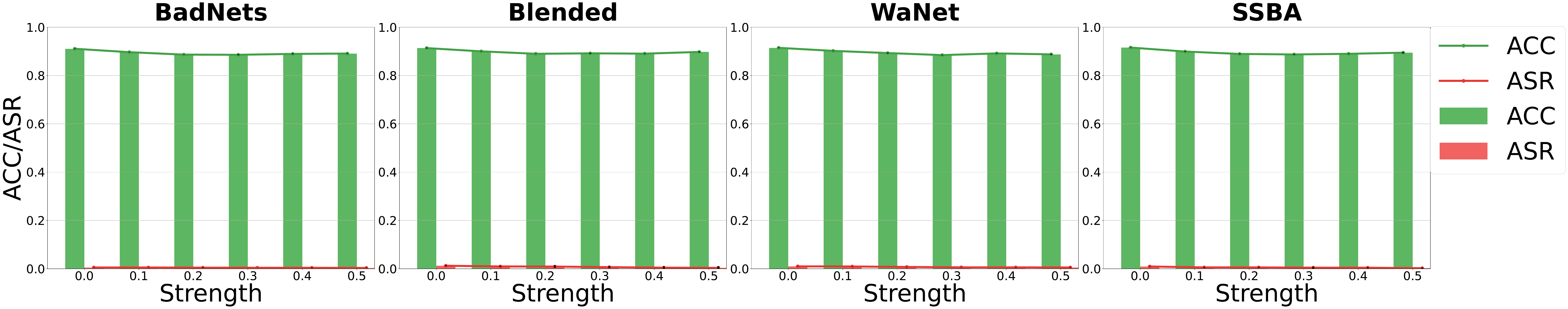}
    \caption{Defense results with different strength of augmentation.}
    \label{fig:noise}
    \end{figure}

In this section, we explore the impact of augmentation on enhancing defensive backdoors using the CIFAR-10 dataset, a 5\% poisoning ratio, and the PreAct-ResNet architecture. We specifically investigate Gaussian noise augmentation by setting $\tau(\vx) = \vx +\alpha * \veps$ with $\veps \sim \mathcal{N}(0, 1)$ determines the augmentation intensity. The results are summarized in Figure~\ref{fig:noise}. Notably, even without any augmentation ($\alpha=0$),  PDB exhibits significant efficacy, likely due to the robustness of the defense mechanisms and our controlled training injection process. Furthermore, as augmentation strength increases, the ASR decreases, albeit at the cost of reduced ACC, indicating a tradeoff between augmentation intensity and model performance.
    
\subsection{Comparison with post-training methods}
Given a reserved dataset, the defender may follow a ‘poisoned first and mitigation later’ manner to train the backdoored model first and employ post-training method to mitigate the backdoor effect.
Therefore, to thoroughly evaluate our method, we compare it with SOTA post-training approaches that aim to remove backdoor effects after model training. To ensure fairness, we adopt the common practice of reserving 5\% of the training data for post-training evaluation. Our experiments are conducted on the CIFAR-10 dataset using PreAct-ResNet18 with a 5\% poisoning ratio. The summarized results in Table~\ref{post} demonstrate that our method consistently achieves superior performance in defending against backdoor attacks, highlighting its promising effectiveness.

\input{table/post}

\subsection{Detailed comparison to NAB}

As previously mentioned, the work most relevant to our research is NAB \cite{liu2023beating}, which presents a promising and impressive approach for enhancing backdoor defense methods. In this context, we conduct a detailed analysis to highlight the distinctions between our method, PDB, and NAB.

\paragraph{PDB does not rely on poisoned sample detection.} Our method diverges from the conventional “detection-mitigation” pipeline by operating independently of any poison detection methods. In contrast, NAB relies on identifying poisoned samples to deploy its defensive trigger.

\paragraph{PDB does not depend on suspicious sample relabeling.} Unlike NAB, our method does not require accurate sample relabeling. In NAB, detected samples are equipped with a defensive trigger and subsequently relabeled. Unfortunately, if a few clean samples are mistakenly labeled as “suspicious,” the defensive trigger may function as a malicious one if further relabeling is incorrect. Note that accurately relabeling a sample is getting harder as the dataset gets larger. Therefore, we observe such scenarios appear more frequently in large datasets during our experiments. For instance, when applying the BadNets attack with a 5\% poisoning ratio on Tiny ImageNet using ViT-B-16, NAB’s model accuracy drops from 73.60\% (without a defensive trigger) to 28.88\% (with the trigger) due to low relabel correctness and poison detection rates.

\subsection{Resistance to ALL2ALL attack}
In this section, we compare PDB with other baselines in ALL2ALL attacks on CIFAR-10 using PreAct-ResNet18. The poisoning ratio is set to 5\% and 10\%. Specifically, the target labels for samples with original labels $y$ are adjusted to $(y+2)\mod K$ (different from the defensive target). The experimental results are summarized in Table~\ref{a2a_cifar10_preactresnet18_05} and Table~\ref{a2a_cifar10_preactresnet18_1}. Notably, PDB achieves the best defending performance, demonstrating superior effectiveness in countering backdoor attacks with multiple targets.

\input{table/a2a_cifar10_preactresnet18_0_05}
\input{table/a2a_cifar10_preactresnet18_0_1}

\subsection{Design of defensive trigger and the satisfaction to Principle 4}
\label{app:p4}
Here, we summarize the aforementioned experiments and provide a comprehensive discussion about \textit{how to meet Principle 4 (Resistance against other backdoors)} from the following perspectives:

\begin{itemize}[leftmargin=*]
    \item \textbf{Design of defensive trigger:}
    \begin{enumerate}
        \item \textbf{Defensive trigger size:} Trigger size directly contributes to the strength of the defensive backdoor. In Table~\ref{reb_size}, we evaluate PDB with a square defensive trigger with sizes ranging from 1x1 to 9x9. From Table~\ref{reb_size}, we can find that \textit{a larger trigger leads to a stronger defensive backdoor}, resulting in a higher ACC and a lower ASR. However, \textit{as the trigger size increases, it may interfere with the visual content of the image}, leading to a slight decrease in ACC. Notably, as the square trigger has strong visibility, a trigger with size 1x1 can still alleviate the malicious backdoor to some extent.
        \input{table/reb_size}

        \item   \textbf{Defensive trigger position:} As aforementioned, the position of the defensive trigger is essential for Principle 3, i.e., minimal impact on model performance. Table~\ref{reb_pos} shows that triggers placed in different positions (corner, random, and center) achieve a similar effect in defending against backdoor attacks. However, \textit{placing a trigger at the center of an image significantly degrades accuracy, as the trigger masks the core patterns of the image.}
        \input{table/reb_pos}

        \item \textbf{Pixel value:} For a square trigger, pixel value is also an important parameter for PDB. In Table~\ref{reb_pix}, we evaluate PDB with a square defensive trigger with different pixel values, from which we can find that \textit{PDB can achieve high effectiveness across different pixel values.}
        \input{table/reb_pix}

    \end{enumerate}

    \item \textbf{Backdoor enhancement strategy during training:}
    \begin{enumerate}
        \item  \textbf{Increasing sampling frequency:} Given a fixed number of defensive poisoned samples, the defensive backdoor can be further enhanced by increasing the sampling frequency of poisoned samples, forcing the model to pay more attention to defensive poisoned samples. Table~\ref{reb_sample} shows that \textit{a larger sampling frequency leads to a stronger defensive backdoor}, resulting in a higher ACC and a lower ASR. Note that for the malicious attacker, the poisoning ratio is expected to be low to ensure the stealthiness of the attack.
        \input{table/reb_sample}
        
        \item \textbf{Data augmentation: } From Fig~\ref{fig:noise}, we can find that PDB, without any sample augmentation ($\alpha=0$), exhibits significant efficacy with ASR lower than 2\%. As augmentation strength increases, the ASR decreases, indicating \textit{a stronger augmentation can help further enhance PDB's effectiveness}. However, a tradeoff between augmentation intensity and model performance is also observed.
    \end{enumerate}
\end{itemize}
In summary, a visible trigger with \textit{larger trigger size}, \textit{higher sampling frequency}, and \textit{data augmentation} contribute to meeting Principle 4.

\subsection{Factors that influence the accuracy of PDB}
Here, we would like to discuss the factors that influence the accuracy of PDB:
\begin{itemize}[leftmargin=*]
    \item \textbf{Model capacity and data complexity:}
    \begin{enumerate}
        \item \textbf{Model capacity:} Since PDB introduces additional task, i.e., injecting defensive backdoor, increasing the model capacity helps to increase the accuracy of PDB, as evidenced in Table~\ref{reb_mod}.
        \input{table/reb_mod}
        \item   \textbf{Dataset complexity:} By comparing defense results with different datasets, we can find that by decreasing the dataset complexity, the accuracy of PDB increases significantly.
    \end{enumerate}
    \item \textbf{Strength of defensive backdoor:}
    \begin{enumerate}
        \item \textbf{Strength of augmentation:} From Figure~\ref{fig:noise}, we can find that there exists a tradeoff between ACC and ASR. Therefore, the accuracy of PDB can be boosted by reducing the strength of augmentation.
        \item  \textbf{Sampling frequency:} From Table~\ref{reb_sample}, we can find that by increasing the sampling frequency of defensive poisoned samples, the accuracy of PDB can be boosted.
        \item \textbf{Trigger size:} Table~\ref{reb_size} shows that a proper choice of trigger size can also help to increase the accuracy. Therefore, if a validation set is accessible, a proper trigger size can be chosen to increase accuracy.
    \end{enumerate}
\end{itemize}
In summary, due to the "home field advantage" of PDB, there are several ways to maintain a high accuracy even in the case of a low malicious poisoning ratio, such as increasing model capacity, simplifying the dataset, reducing the strength of augmentation to defensive poisoned samples, increasing the sampling frequency and choosing a proper defensive trigger size.
  
\subsection{ Novelty and comparison with backdoorIndicator}

Here, we highlight the distinctions between our approach and BackdoorIndicator \cite{li2024backdoorindicator}.

\textbf{First}, we would like to clarify the following differences between our work and backdoorIndicator \cite{li2024backdoorindicator}:
\begin{itemize}[leftmargin=*]
  \item \textbf{Threat model:} \cite{li2024backdoorindicator} targets decentralized training (FL) setting, where multiple clients train models locally and contribute updates to a central server. Our work considers a centralized training setting where only a central server is used.
  \item \textbf{Task:} \cite{li2024backdoorindicator} focuses on detecting malicious clients, whereas our method aims to train a secure model on a poisoned dataset without clients.
  \item \textbf{Motivation:} \cite{li2024backdoorindicator} is built on the motivation that planting subsequent backdoors with the same target label enhances previously planted backdoors, therefore, providing a way to detect the poisoned clients, while our method is based on the motivation that planting a concurrent reversible backdoor can help to mitigate the malicious backdoor.
  \item \textbf{Methodology:} \cite{li2024backdoorindicator} utilizes OOD samples for backdoor client detection while our method constructs a proactive defensive poison dataset, following well-designed principles.
\end{itemize}

\textbf{Second}, we would like to discuss the challenges in direct utilizing backdoorIndicator in our setting:
\begin{itemize}[leftmargin=*]
  \item BackdoorIndicator is designed to detect malicious clients within a federated learning (FL) context. This makes it challenging to apply BackdoorIndicator directly to our centralized environment since the task of identifying backdoored clients does not naturally fit into this setting (only a central server).
  \item For comparison between BackdoorIndicator and our method, we need to emulate an FL scenario by assigning each image to a separate client (ensuring existence of benign client), thereby creating 50,000 local models from the CIFAR-10 dataset to defend a single attack with PreAct-ResNet18. This would require an impractical amount of computational resources, estimated at over 30,000 hours (1,250 days) of training time and 30TB of storage space using a server with a single RTX 3090 GPU.
\end{itemize}

\subsection{Comparison to FT-SAM}
To address the comparison with FT-SAM \cite{Zhu_2023_ICCV}, we have adapted their method to our experimental setting. It's worth noting that in  \cite{Zhu_2023_ICCV}, the authors employ the Blended attack with a blending ratio of 0.1 (Blended-0.1), whereas we use a blending ratio of 0.2 (Blended-0.2). For consistency and completeness, we have now included experiments using both blending ratios, and the results are shown below:

\input{table/reb_sam}

From Table~\ref{reb_sam}, we can find FT-SAM can achieve a higher accuracy as it aims to fine-tune a backdoored model while PDB aims to train a model from scratch. Consistent with \cite{Zhu_2023_ICCV}, Table~\ref{reb_sam} shows that FT-SAM can mitigate backdoor attacks for most cases, except for the Blended-0.2. We observe that FT-SAM struggles to defend against blended attacks with higher blending ratios, such as 0.2. Notably, PDB achieves a significantly lower ASR across all cases, with an average ASR below 0.5\%.

%% file: table/fake.tex
\begin{table}[H]
\centering
\caption{Defense results using generated dataset on CIFAR-10 and PreAct-ResNet18 (\%).}
\label{fake}
\setlength{\tabcolsep}{3pt} 
\renewcommand{\arraystretch}{1.5} 
\scalebox{0.62}{
\begin{tabular}{@{}c|cc|cc|cc|cc|cc@{}}
\toprule
Attack $\rightarrow$              & \multicolumn{2}{c|}{BadNets \cite{gu2019badnets}} & \multicolumn{2}{c|}{Blended \cite{chen2017targeted}} & \multicolumn{2}{c|}{SIG \cite{barni2019new}} & \multicolumn{2}{c|}{SSBA \cite{li2021invisible}} & \multicolumn{2}{c|}{Average} \\ \midrule
Defense $\downarrow$              & ACC          & ASR          & ACC          & ASR          & ACC        & ASR        & ACC         & ASR        & ACC          & ASR          \\  \midrule
No Defense           & 92.64        & 88.74        & 93.67        & 99.61        & 93.64      & 97.09      & 93.27       & 94.91      & 93.31        & 95.09        \\
PDB (\textbf{Ours})                  & 91.08        & 0.38         & 91.36        & 0.70         & 91.79      & 0.06       & 91.58       & 0.46       & 91.45        & 0.40         \\
PDB (\textbf{Ours})+Generate Dataset & 90.17        & 0.48         & 90.90        & 0.97         & 91.33      & 0.00       & 91.49       & 0.56       & 91.24        & 0.78   \\ \bottomrule
\end{tabular}}
\end{table}

%% file: table/config.tex
\begin{table}[H]
\centering
\caption{Results on PDB with different configurations.}
\label{config}
\setlength{\tabcolsep}{3pt} 
\renewcommand{\arraystretch}{1.5} 
\scalebox{0.62}{
\begin{tabular}{@{}c|cc|cc|cc|cc|cc@{}}
\toprule
\multicolumn{3}{c|}{Configuration $\rightarrow$}          & \multicolumn{2}{c|}{Config 1 (a+$h_1$)} & \multicolumn{2}{c|}{Config 2 (b+$h_1$)} & \multicolumn{2}{c|}{Config 3 (c+$h_1$)} & \multicolumn{2}{c}{Config 4 (a+$h_2$)} \\ \midrule
Defense $\rightarrow$ & \multicolumn{2}{c|}{No Defense} & \multicolumn{2}{c|}{PDB (\textbf{Ours})}                         & \multicolumn{2}{c|}{PDB (\textbf{Ours})}                & \multicolumn{2}{c|}{PDB (\textbf{Ours})}                & \multicolumn{2}{c}{PDB (\textbf{Ours})}                \\ \midrule
Attack $\downarrow$ & ACC            & ASR           & ACC                         & ASR               & ACC            & ASR                   & ACC                & ASR               & ACC            & ASR                   \\ \midrule
BadNets \cite{gu2019badnets}  & 92.64          & 88.74         & \textbf{91.08}          & \textbf{0.38}         & 90.09              & 0.40              & 90.89                   & 0.76         & 89.93              & 0.81              \\
Blended \cite{chen2017targeted} & 93.67          & 99.61         & 91.36                   & 0.70                  & 89.20              & 0.04              & 90.45                   & 0.03         & \textbf{91.96}     & \textbf{0.00}     \\
SIG \cite{barni2019new}     & 93.64          & 97.09         & 91.79                   & 0.06                  & 91.17              & 0.10              & \textbf{92.13}          & 0.38         & 91.51              & \textbf{0.00}     \\
SSBA \cite{li2021invisible}    & 93.27          & 94.91         & 91.58                   & 0.46                  & 91.03              & 0.21              & 89.58                   & 0.37         & \textbf{91.85}     & \textbf{0.09}     \\
Average & 93.31          & 95.09         & \textbf{91.45}          & 0.40                  & 90.37              & 0.19              & 90.76                   & 0.38         & 91.31              & \textbf{0.23}    \\ \bottomrule
\end{tabular}}
\end{table}

%% file: table/a2a_1.tex
\begin{table}[H]
\centering
\caption{Results on attacks with the same target label as defensive backdoor.}
\label{a2a_1}
\setlength{\tabcolsep}{3pt} 
\renewcommand{\arraystretch}{1.5} 
\scalebox{0.62}{
\begin{tabular}{@{}c|cc|cc|cc|cc@{}}
\toprule
Poisoning Ratio $\rightarrow$ & \multicolumn{4}{c}{5\%}                                  & \multicolumn{4}{c}{10\%}                                 \\ \midrule
Defense  $\rightarrow$         & \multicolumn{2}{c}{No Defense} & \multicolumn{2}{c}{PDB (\textbf{Ours})} & \multicolumn{2}{c}{No Defense} & \multicolumn{2}{c}{PDB (\textbf{Ours})} \\ \midrule
Attack  $\downarrow$        & ACC            & ASR           & ACC         & ASR       & ACC            & ASR           & ACC         & ASR       \\ \midrule
BadNet          & 92.54          & 65.85         & 91.37       & 1.01      & 91.89          & 74.42         & 91.26       & 1.00      \\
Blended \cite{chen2017targeted}         & 93.71          & 83.68         & 92.28       & 1.64      & 93.74          & 86.85         & 92.03       & 1.99      \\
BPP \cite{Wang_2022_CVPR}             & 90.92          & 86.42         & 91.89       & 1.45      & 91.49          & 87.71         & 91.95       & 1.57      \\
SIG \cite{barni2019new}             & 93.73          & 88.02         & 92.02       & 1.59      & 93.40          & 90.61         & 91.91       & 1.29      \\
Average         & 92.73          & 80.99         & 91.89       & 1.42      & 92.63          & 84.90         & 91.79       & 1.46     \\
\bottomrule
\end{tabular}}
\end{table}

%% file: table/post.tex
\begin{table}[H]
\centering
\caption{Results on CIFAR-10 with PreAct-ResNet18 and poisoning ratio $5.0\%$.}
\label{post}
\setlength{\tabcolsep}{3pt} 
\renewcommand{\arraystretch}{1.5} 
\scalebox{0.62}{
\begin{tabular}{c|cc|ccc|ccc|ccc|ccc|ccc|ccc}
\toprule
Defense $\rightarrow$ & \multicolumn{2}{c|}{No Defense} & \multicolumn{3}{c|}{FT} & \multicolumn{3}{c|}{FP \cite{liu2018fine}} & \multicolumn{3}{c|}{NC \cite{wang2019neural}} & \multicolumn{3}{c|}{NAD \cite{li2021neural}} & \multicolumn{3}{c|}{i-BAU \cite{zeng2022adversarial}} & \multicolumn{3}{c}{PDB (\textbf{Ours})} \\ \midrule
Attack $\downarrow$ & \multicolumn{1}{c}{ACC} & \multicolumn{1}{c|}{ASR} & \multicolumn{1}{c}{ACC} & \multicolumn{1}{c}{ASR} & \multicolumn{1}{c|}{DER} & \multicolumn{1}{c}{ACC} & \multicolumn{1}{c}{ASR} & \multicolumn{1}{c|}{DER} & \multicolumn{1}{c}{ACC} & \multicolumn{1}{c}{ASR} & \multicolumn{1}{c|}{DER} & \multicolumn{1}{c}{ACC} & \multicolumn{1}{c}{ASR} & \multicolumn{1}{c|}{DER} & \multicolumn{1}{c}{ACC} & \multicolumn{1}{c}{ASR} & \multicolumn{1}{c|}{DER} & \multicolumn{1}{c}{ACC} & \multicolumn{1}{c}{ASR} & \multicolumn{1}{c}{DER} \\
\midrule
BadNets \cite{gu2019badnets} & $92.64$& $88.74$& $\underline{91.22}$& $4.49$& $91.42$& $\textbf{92.47}$& $17.47$& $85.55$& $89.89$& $1.17$& $\textbf{92.41}$& $91.03$& $4.73$& $91.20$& $89.21$& $\underline{0.81}$& $92.25$& $89.34$& $\textbf{0.66}$& $\underline{92.39}$\\
Blended \cite{chen2017targeted} & $93.67$& $99.61$& $93.25$& $98.78$& $50.21$& $\underline{93.46}$& $98.87$& $50.27$& $\textbf{93.66}$& $99.61$& $50.00$& $93.10$& $99.06$& $49.99$& $85.92$& $\underline{36.33}$& $\underline{77.76}$& $90.24$& $\textbf{0.87}$& $\textbf{97.66}$\\
SIG \cite{barni2019new} & $93.64$& $97.09$& $92.84$& $96.42$& $49.93$& $\underline{93.27}$& $99.79$& $49.81$& $\textbf{93.65}$& $97.09$& $50.00$& $92.49$& $96.98$& $49.48$& $86.12$& $\underline{3.50}$& $\underline{93.03}$& $90.26$& $\textbf{0.01}$& $\textbf{96.85}$\\
SSBA \cite{li2021invisible} & $93.27$& $94.91$& $\textbf{93.08}$& $87.46$& $53.63$& $\underline{93.05}$& $87.14$& $53.77$& $91.04$& $\underline{0.80}$& $\textbf{95.94}$& $92.49$& $88.63$& $52.75$& $89.43$& $2.30$& $94.39$& $90.02$& $\textbf{0.38}$& $\underline{95.64}$\\
WaNet \cite{nguyen2021wanet} & $91.76$& $85.5$& $\textbf{93.47}$& $31.32$& $77.09$& $92.14$& $26.10$& $79.7$& $91.76$& $85.50$& $50.00$& $\underline{93.31}$& $50.40$& $67.55$& $91.13$& $\underline{6.11}$& $\underline{89.38}$& $89.96$& $\textbf{0.90}$& $\textbf{91.40}$\\
BPP \cite{Wang_2022_CVPR} & $91.47$& $99.34$& $\textbf{93.37}$& $3.46$& $\textbf{97.94}$& $85.98$& $\textbf{3.11}$& $95.37$& $91.47$& $99.34$& $50.00$& $\underline{93.09}$& $3.53$& $\underline{97.91}$& $91.35$& $5.72$& $96.75$& $90.9$& $\underline{3.17}$& $97.80$\\
Trojan \cite{Trojannn} & $93.79$& $99.99$& $\textbf{92.90}$& $44.22$& $77.44$& $83.89$& $12.11$& $88.99$& $92.59$& $95.06$& $51.87$& $\underline{92.68}$& $\underline{4.24}$& $\underline{97.32}$& $89.42$& $7.49$& $94.06$& $89.64$& $\textbf{0.19}$& $\textbf{97.82}$\\
Average & $92.89$& $95.03$& $\textbf{92.88}$& $52.31$& $71.09$& $90.61$& $49.23$& $71.92$& $92.01$& $68.37$& $62.89$& $\underline{92.60}$& $49.65$& $72.31$& $88.94$& $\underline{8.90}$& $\underline{91.09}$& $90.05$& $\textbf{0.88}$& $\textbf{95.65}$\\

\bottomrule

\end{tabular}
}
\end{table}

%% file: table/a2a_cifar10_preactresnet18_0_05.tex
\begin{table}[H]
\centering
\caption{ALL2ALL attack results on CIFAR-10 with PreAct-ResNet18 and poisoning ratio $5.0\%$.}
\label{a2a_cifar10_preactresnet18_05}
\setlength{\tabcolsep}{3pt} 
\renewcommand{\arraystretch}{1.5} 
\scalebox{0.62}{
\begin{tabular}{c|cc|ccc|ccc|ccc|ccc|ccc|ccc}
\toprule
Defense $\rightarrow$ & \multicolumn{2}{c|}{No Defense} & \multicolumn{3}{c|}{AC \cite{chen2019detecting}} & \multicolumn{3}{c|}{Spectral \cite{tran2018spectral}} & \multicolumn{3}{c|}{ABL \cite{li2021anti}} & \multicolumn{3}{c|}{DBD \cite{huang2022backdoor}} & \multicolumn{3}{c|}{NAB \cite{liu2023beating}} & \multicolumn{3}{c}{PDB (\textbf{Ours})} \\ \midrule
Attack $\downarrow$ & \multicolumn{1}{c}{ACC} & \multicolumn{1}{c|}{ASR} & \multicolumn{1}{c}{ACC} & \multicolumn{1}{c}{ASR} & \multicolumn{1}{c|}{DER} & \multicolumn{1}{c}{ACC} & \multicolumn{1}{c}{ASR} & \multicolumn{1}{c|}{DER} & \multicolumn{1}{c}{ACC} & \multicolumn{1}{c}{ASR} & \multicolumn{1}{c|}{DER} & \multicolumn{1}{c}{ACC} & \multicolumn{1}{c}{ASR} & \multicolumn{1}{c|}{DER} & \multicolumn{1}{c}{ACC} & \multicolumn{1}{c}{ASR} & \multicolumn{1}{c|}{DER} & \multicolumn{1}{c}{ACC} & \multicolumn{1}{c}{ASR} & \multicolumn{1}{c}{DER} \\
\midrule
BadNets \cite{gu2019badnets} & $92.5$& $61.33$& $90.1$& $53.7$& $52.61$& $\textbf{92.33}$& $57.73$& $51.72$& $52.46$& $59.96$& $30.66$& $87.1$& $\underline{4.52}$& $\underline{75.70}$& $80.51$& $62.74$& $44.0$& $\underline{90.68}$& $\textbf{2.72}$& $\textbf{78.40}$\\
Blended \cite{chen2017targeted} & $93.51$& $83.87$& $91.36$& $78.56$& $51.58$& $\textbf{93.72}$& $84.66$& $50.0$& $68.04$& $35.62$& $61.39$& $75.24$& $\underline{26.62}$& $\underline{69.49}$& $90.34$& $79.09$& $50.8$& $\underline{91.87}$& $\textbf{3.95}$& $\textbf{89.14}$\\
SIG \cite{barni2019new} & $93.52$& $88.15$& $91.49$& $83.07$& $51.52$& $\textbf{94.02}$& $88.77$& $50.0$& $67.2$& $59.67$& $51.08$& $76.19$& $\underline{20.26}$& $\underline{75.28}$& $82.65$& $83.19$& $47.04$& $\underline{91.73}$& $\textbf{3.13}$& $\textbf{91.62}$\\
BPP \cite{Wang_2022_CVPR} & $90.92$& $86.42$& $91.26$& $83.67$& $51.37$& $\textbf{93.72}$& $87.52$& $50.0$& $33.44$& $19.69$& $54.62$& $77.48$& $\underline{2.24}$& $\underline{85.37}$& $86.32$& $83.33$& $49.24$& $\underline{91.89}$& $\textbf{1.45}$& $\textbf{92.48}$\\
Average & $92.61$& $79.94$& $91.05$& $74.75$& $51.77$& $\textbf{93.45}$& $79.67$& $50.43$& $55.28$& $43.74$& $49.44$& $79.0$& $\underline{13.41}$& $\underline{76.46}$& $84.96$& $77.09$& $47.77$& $\underline{91.54}$& $\textbf{2.81}$& $\textbf{87.91}$\\

\bottomrule

\end{tabular}
}
\end{table}

%% file: table/a2a_cifar10_preactresnet18_0_1.tex
\begin{table}[H]
\centering
\caption{ALL2ALL attack results on CIFAR-10 with PreAct-ResNet18 and poisoning ratio $10.0\%$.}
\label{a2a_cifar10_preactresnet18_1}
\setlength{\tabcolsep}{3pt} 
\renewcommand{\arraystretch}{1.5} 
\scalebox{0.62}{
\begin{tabular}{c|cc|ccc|ccc|ccc|ccc|ccc|ccc}
\toprule
Defense $\rightarrow$ & \multicolumn{2}{c|}{No Defense} & \multicolumn{3}{c|}{AC \cite{chen2019detecting}} & \multicolumn{3}{c|}{Spectral \cite{tran2018spectral}} & \multicolumn{3}{c|}{ABL \cite{li2021anti}} & \multicolumn{3}{c|}{DBD \cite{huang2022backdoor}} & \multicolumn{3}{c|}{NAB \cite{liu2023beating}} & \multicolumn{3}{c}{PDB (\textbf{Ours})} \\ \midrule
Attack $\downarrow$ & \multicolumn{1}{c}{ACC} & \multicolumn{1}{c|}{ASR} & \multicolumn{1}{c}{ACC} & \multicolumn{1}{c}{ASR} & \multicolumn{1}{c|}{DER} & \multicolumn{1}{c}{ACC} & \multicolumn{1}{c}{ASR} & \multicolumn{1}{c|}{DER} & \multicolumn{1}{c}{ACC} & \multicolumn{1}{c}{ASR} & \multicolumn{1}{c|}{DER} & \multicolumn{1}{c}{ACC} & \multicolumn{1}{c}{ASR} & \multicolumn{1}{c|}{DER} & \multicolumn{1}{c}{ACC} & \multicolumn{1}{c}{ASR} & \multicolumn{1}{c|}{DER} & \multicolumn{1}{c}{ACC} & \multicolumn{1}{c}{ASR} & \multicolumn{1}{c}{DER} \\
\midrule
BadNets \cite{gu2019badnets} & $91.82$& $72.2$& $88.89$& $65.73$& $51.77$& $\textbf{90.54}$& $72.11$& $49.4$& $26.34$& $27.9$& $39.41$& $85.52$& $\underline{5.25}$& $\underline{80.32}$& $83.47$& $60.34$& $51.76$& $\underline{90.53}$& $\textbf{2.49}$& $\textbf{84.21}$\\
Blended \cite{chen2017targeted} & $93.54$& $87.04$& $91.01$& $82.12$& $51.19$& $\textbf{93.83}$& $87.09$& $50.0$& $74.11$& $56.56$& $55.53$& $80.08$& $\underline{22.48}$& $\underline{75.55}$& $88.68$& $79.7$& $51.24$& $\underline{91.89}$& $\textbf{3.02}$& $\textbf{91.18}$\\
SIG \cite{barni2019new} & $93.64$& $90.65$& $91.18$& $87.14$& $50.52$& $\textbf{93.58}$& $90.68$& $49.97$& $86.47$& $82.69$& $50.4$& $68.23$& $\underline{42.20}$& $\underline{61.52}$& $82.05$& $83.85$& $47.6$& $\underline{92.08}$& $\textbf{2.68}$& $\textbf{93.20}$\\
BPP \cite{Wang_2022_CVPR} & $91.49$& $87.71$& $91.66$& $85.97$& $50.87$& $\textbf{93.99}$& $89.41$& $50.0$& $75.58$& $74.15$& $48.82$& $81.9$& $\underline{2.68}$& $\underline{87.72}$& $84.44$& $81.52$& $49.57$& $\underline{91.95}$& $\textbf{1.57}$& $\textbf{93.07}$\\
Average & $92.62$& $84.4$& $90.68$& $80.24$& $51.09$& $\textbf{92.98}$& $84.82$& $49.84$& $65.62$& $60.32$& $48.54$& $78.93$& $\underline{18.15}$& $\underline{76.28}$& $84.66$& $76.35$& $50.04$& $\underline{91.61}$& $\textbf{2.44}$& $\textbf{90.42}$\\

\bottomrule

\end{tabular}
}
\end{table}

%% file: table/reb_size.tex
\begin{table}[!ht]
    \centering
    \caption{Results on PreAct-ResNet18 with Poisoning Ratio 5\% and different defensive trigger size}
    \label{reb_size}
    \setlength{\tabcolsep}{3pt} 
\renewcommand{\arraystretch}{1.5} 
\scalebox{0.62}{
    \begin{tabular}{c|c|c|c|c|c|c|c|c|c|c}
    \toprule
        Attack → & BadNet & BadNet & Blended & Blended & Sig & Sig & SSBA & SSBA & WaNet & WaNet \\ \midrule
        Defensive Trigger size ↓ & ACC & ASR & ACC & ASR & ACC & ASR & ACC & ASR & ACC & ASR \\ \midrule
        1x1 & 48.75 & 6.88 & 52.34 & 5.06 & 53.15 & 5.22 & 52.03 & 6.20 & 58.04 & 4.26 \\ 
        2x2 & 74.39 & 3.37 & 81.38 & 2.71 & 81.13 & 2.32 & 77.49 & 3.87 & 76.49 & 3.98 \\ 
        3x3 & 86.08 & 0.26 & 85.60 & 0.67 & 86.94 & 0.07 & 87.01 & 0.49 & 85.70 & 0.97 \\ 
        4x4 & 89.46 & 0.28 & 90.07 & 0.56 & 90.38 & 0.07 & 89.60 & 0.44 & 89.98 & 0.92 \\ 
        5x5 & 91.51 & 0.33 & 92.22 & 0.31 & 92.35 & 0.06 & 92.14 & 0.64 & 92.05 & 0.97 \\ 
        6x6 & 90.78 & 0.32 & 91.93 & 0.49 & 92.04 & 0.04 & 91.82 & 0.41 & 91.52 & 0.91 \\ 
        7x7 & 91.08 & 0.38 & 91.36 & 0.70 & 91.79 & 0.06 & 91.58 & 0.46 & 91.47 & 0.92 \\ 
        8x8 & 90.48 & 0.33 & 91.56 & 0.40 & 91.59 & 0.02 & 91.41 & 0.39 & 91.44 & 0.86 \\ 
        9x9 & 90.21 & 0.32 & 91.24 & 0.39 & 90.79 & 0.03 & 90.92 & 0.32 & 90.87 & 0.56 \\ 
        \bottomrule
    \end{tabular}}
\end{table}

%% file: table/reb_pos.tex
\begin{table}[!ht]
    \centering
    \caption{Results on PreAct-ResNet18 with Poisoning Ratio 5\% and different positions}
    \label{reb_pos}
    \setlength{\tabcolsep}{3pt} 
\renewcommand{\arraystretch}{1.5} 
\scalebox{0.62}{
    \begin{tabular}{c|c|c|c|c|c|c|c|c|c|c}
        \toprule
        Attack → & BadNet & BadNet & Blended & Blended & Sig & Sig & SSBA & SSBA & WaNet & WaNet \\ \midrule
        Defensive Trigger Position ↓ & ACC & ASR & ACC & ASR & ACC & ASR & ACC & ASR & ACC & ASR \\ \midrule
        Corner & 91.08 & 0.38 & 91.36 & 0.7 & 91.79 & 0.06 & 91.58 & 0.46 & 91.47 & 0.92 \\ 
        Random & 88.79 & 0.69 & 90.10 & 0.81 & 90.12 & 0.16 & 89.39 & 0.66 & 89.49 & 0.97 \\ 
        Center & 87.35 & 0.63 & 87.82 & 0.44 & 88.19 & 0.06 & 87.93 & 0.89 & 87.70 & 0.93 \\ \bottomrule
    \end{tabular}}
\end{table}

%% file: table/reb_pix.tex
\begin{table}[!ht]
    \centering
    \caption{Results on PreAct-ResNet18 with Poisoning Ratio 5\% and different pixel values}
    \label{reb_pix}
    \setlength{\tabcolsep}{3pt} 
\renewcommand{\arraystretch}{1.5} 
\scalebox{0.62}{
    \begin{tabular}{c|c|c|c|c|c|c|c|c|c|c}
    \toprule
        Attack → & BadNet & BadNet & Blended & Blended & Sig & Sig & SSBA & SSBA & WaNet & WaNet \\ \midrule
        Pixel ↓ & ACC & ASR & ACC & ASR & ACC & ASR & ACC & ASR & ACC & ASR \\ \midrule
        1.50 & 90.69 & 0.57 & 91.40 & 0.56 & 91.74 & 0.09 & 91.54 & 0.60 & 91.44 & 0.83 \\ 
        2.00 & 91.08 & 0.38 & 91.36 & 0.7 & 91.79 & 0.06 & 91.58 & 0.46 & 91.47 & 0.92 \\ 
        2.50 & 90.99 & 0.48 & 91.39 & 0.50 & 91.77 & 0.04 & 91.48 & 0.80 & 91.54 & 0.54 \\ 
        -0.50 & 90.94 & 0.48 & 91.78 & 0.91 & 91.56 & 0.01 & 91.64 & 0.61 & 91.69 & 0.60 \\ 
        -1.00 & 90.84 & 0.23 & 92.31 & 0.07 & 91.81 & 0.00 & 91.85 & 1.07 & 91.83 & 0.90 \\ 
        -1.50 & 90.86 & 0.29 & 91.62 & 1.00 & 91.88 & 0.00 & 91.43 & 0.66 & 91.86 & 0.78 \\
        \bottomrule
    \end{tabular}}
\end{table}

%% file: table/reb_sample.tex
\begin{table}[!ht]
    \centering
    \caption{Results on PreAct-ResNet18 with poisoning ratio 5\% and different sampling frequencies}
    \label{reb_sample}
    \setlength{\tabcolsep}{3pt} 
\renewcommand{\arraystretch}{1.5} 
\scalebox{0.62}{
    \begin{tabular}{c|c|c|c|c|c|c|c|c|c|c}
        \toprule
        Attack → & BadNet & BadNet & Blended & Blended & Sig & Sig & SSBA & SSBA & WaNet & WaNet \\ \midrule
        Frequency ↓ & ACC & ASR & ACC & ASR & ACC & ASR & ACC & ASR & ACC & ASR \\ \midrule
        1 & 91.01 & 0.69 & 91.19 & 1.48 & 91.38 & 4.62 & 91.19 & 1.24 & 91.13 & 1.07 \\ 
        3 & 91.06 & 0.57 & 91.27 & 1.39 & 91.73 & 0.10 & 91.28 & 0.72 & 91.44 & 0.97 \\ 
        5 & 91.08 & 0.38 & 91.36 & 0.70 & 91.79 & 0.06 & 91.58 & 0.46 & 91.47 & 0.92 \\ 
        7 & 91.34 & 0.27 & 91.56 & 0.59 & 91.98 & 0.04 & 91.89 & 0.43 & 91.84 & 0.27 \\ 
        9 & 92.15 & 0.20 & 92.19 & 0.50 & 92.27 & 0.02 & 92.30 & 0.31 & 92.48 & 0.16 \\ \bottomrule
    \end{tabular}}
\end{table}

%% file: table/reb_mod.tex
\begin{table}[!ht]
    \centering
    \caption{Results on different models}
    \label{reb_mod}
    \setlength{\tabcolsep}{3pt} 
\renewcommand{\arraystretch}{1.5} 
\scalebox{0.62}{
    \begin{tabular}{c|c|c|c|c|c|c}
        \toprule
        Model & ResNet-18 & ResNet-18 & ResNet-34 & ResNet-34 & ResNet-50 & ResNet-50 \\ \midrule
        Metric & ACC & ASR & ACC & ASR & ACC & ASR \\ \midrule
        No Defense & 92.54 & 76.27 & 93.08 & 82.48 & 93.76 & 87.26 \\ 
        PDB & 91.81 & 0.29 & 92.63 & 0.28 & 93.67 & 0.18 \\ \bottomrule
    \end{tabular}}
\end{table}

%% file: table/reb_sam.tex
\begin{table}[!ht]
    \centering
        \caption{Results on PreAct-ResNet18 with FT-SAM}
    \label{reb_sam}
    \setlength{\tabcolsep}{3pt} 
\renewcommand{\arraystretch}{1.5} 
\scalebox{0.62}{
    \begin{tabular}{c|c|c|c|c|c|c|c|c|c|c|c}
        \toprule
        ~ & Attack → & BadNet & BadNet & Blended-0.2 & Blended-0.2 & Blended-0.1 & Blended-0.1 & Sig & Sig & SSBA & SSBA \\ \midrule
        Poisoning ratio ↓ & Defense ↓ & ACC & ASR & ACC & ASR & ACC & ASR & ACC & ASR & ACC & ASR \\ \midrule
        5\% & FT-SAM & 92.66 & 1.22 & 92.87 & 31.54 & 92.76 & 2.87 & 92.82 & 1.80 & 92.83 & 3.27 \\ 
        5\% & PDB & 91.08 & 0.38 & 91.36 & 0.70 & 91.85 & 0.22 & 91.79 & 0.06 & 91.58 & 0.46 \\ \bottomrule
    \end{tabular}}
\end{table}

%% file: section/checklist.tex
\section*{NeurIPS Paper Checklist}

\begin{enumerate}

\item {\bf Claims}
    \item[] Question: Do the main claims made in the abstract and introduction accurately reflect the paper's contributions and scope?
    \item[] Answer: \answerYes{} 
    \item[] Justification: The abstract and introduction of the paper accurately reflect the paper's contributions and scope. They clearly state that the paper addresses the challenge of training a clean model on a potentially poisoned dataset by proposing a novel defense mechanism.

\item {\bf Limitations}
    \item[] Question: Does the paper discuss the limitations of the work performed by the authors?
    \item[] Answer: \answerYes{} 
    \item[] Justification: The paper discusses the limitations of the work performed by the authors. In the "Limitations and future work" section, we acknowledge that their implementation of the reversible backdoor defense has several key limitations.

\item {\bf Theory Assumptions and Proofs}
    \item[] Question: For each theoretical result, does the paper provide the full set of assumptions and a complete (and correct) proof?
    \item[] Answer: \answerNA{} 
    \item[] Justification: The paper does not include theoretical results. 

    \item {\bf Experimental Result Reproducibility}
    \item[] Question: Does the paper fully disclose all the information needed to reproduce the main experimental results of the paper to the extent that it affects the main claims and/or conclusions of the paper (regardless of whether the code and data are provided or not)?
    \item[] Answer: \answerYes{} 
    \item[] Justification: The paper appears to provide detailed information on the experimental setup based on a popular Benchmark Project, including the datasets used, the model architectures, the backdoor attack strategies, and the defense mechanisms compared.
    

\item {\bf Open access to data and code}
    \item[] Question: Does the paper provide open access to the data and code, with sufficient instructions to faithfully reproduce the main experimental results, as described in supplemental material?
    \item[] Answer: \answerYes{} 
    \item[] Justification: Provided in the supplemental material.

\item {\bf Experimental Setting/Details}
    \item[] Question: Does the paper specify all the training and test details (e.g., data splits, hyperparameters, how they were chosen, type of optimizer, etc.) necessary to understand the results?
    \item[] Answer: \answerYes{} 
    \item[] Justification: Provided in the main text and the appendix.

\item {\bf Experiment Statistical Significance}
    \item[] Question: Does the paper report error bars suitably and correctly defined or other appropriate information about the statistical significance of the experiments?
    \item[] Answer: \answerYes{} 
    \item[] Justification: Provided in the main text and the appendix. Due to space limit, the bar are mainly visualized in the plots.

\item {\bf Experiments Compute Resources}
    \item[] Question: For each experiment, does the paper provide sufficient information on the computer resources (type of compute workers, memory, time of execution) needed to reproduce the experiments?
    \item[] Answer: \answerYes{} 
    \item[] Justification: Provided in the main text and the appendix.
    
\item {\bf Code Of Ethics}
    \item[] Question: Does the research conducted in the paper conform, in every respect, with the NeurIPS Code of Ethics \url{https://neurips.cc/public/EthicsGuidelines}?
    \item[] Answer: \answerYes{} 
    \item[] Justification: 

\item {\bf Broader Impacts}
    \item[] Question: Does the paper discuss both potential positive societal impacts and negative societal impacts of the work performed?
    \item[] Answer: \answerYes{} 
    \item[] Justification: Provided in the main text.
    
\item {\bf Safeguards}
    \item[] Question: Does the paper describe safeguards that have been put in place for responsible release of data or models that have a high risk for misuse (e.g., pretrained language models, image generators, or scraped datasets)?
    \item[] Answer: \answerNA{} 
    \item[] Justification: No such risks.

\item {\bf Licenses for existing assets}
    \item[] Question: Are the creators or original owners of assets (e.g., code, data, models), used in the paper, properly credited and are the license and terms of use explicitly mentioned and properly respected?
    \item[] Answer: \answerYes{} 
    \item[] Justification: Properly credited.

\item {\bf New Assets}
    \item[] Question: Are new assets introduced in the paper well documented and is the documentation provided alongside the assets?
    \item[] Answer: \answerNA{} 
    \item[] Justification: Do not release new assets.

\item {\bf Crowdsourcing and Research with Human Subjects}
    \item[] Question: For crowdsourcing experiments and research with human subjects, does the paper include the full text of instructions given to participants and screenshots, if applicable, as well as details about compensation (if any)? 
    \item[] Answer: \answerNA{} 
    \item[] Justification: Not involve crowdsourcing nor research with human subjects.

\item {\bf Institutional Review Board (IRB) Approvals or Equivalent for Research with Human Subjects}
    \item[] Question: Does the paper describe potential risks incurred by study participants, whether such risks were disclosed to the subjects, and whether Institutional Review Board (IRB) approvals (or an equivalent approval/review based on the requirements of your country or institution) were obtained?
    \item[] Answer: \answerNA{} 
    \item[] Justification: Not involve crowdsourcing nor research with human subjects.

\end{enumerate}